\documentclass[manuscript]{aastex62}

\usepackage{amsmath}
\usepackage{bm}
\usepackage{url}
\usepackage{amssymb}
\usepackage{amsmath}
\usepackage{graphicx}
\usepackage{subfigure}
\usepackage{tabls}
\usepackage{ctable}
\usepackage{booktabs}
\usepackage{longtable}
\usepackage{rotating}
\usepackage{xcolor}
\usepackage{multirow}
\usepackage{rotating,tabularx}
\usepackage{listings}
\usepackage{color}
\usepackage{colortbl}
\usepackage{hhline}

\begin{document}

\title{Do Cellular Automaton Avalanche Models Simulate the Quasi-Periodic Pulsations of Solar Flares?}

\correspondingauthor{Nastaran Farhang}
\email{farhangnastaran@gmail.com}
\email{safari@znu.ac.ir}

\author{Nastaran Farhang}
\affil{Department of Physics, Isfahan University of Technology, 84156-83111, Isfahan, Iran}

\author{Farhad Shahbazi}
\affil{Department of Physics, Isfahan University of Technology, 84156-83111, Isfahan, Iran}

\author{Hossein Safari}
\affiliation{Department of Physics, Faculty of Science, University of Zanjan, 45195-313, Zanjan, Iran}

\begin{abstract}
Quasi-periodic pulsations (QPPs) with various periods that originate in the underlying magnetohydrodynamic processes of the flaring structures are detected repeatedly in the solar flare emissions. We apply a 2D cellular automaton (CA) avalanche model to simulate QPPs as a result of a repetitive load/unload mechanism. We show that the frequent occurrence of magnetic reconnections in a flaring loop could induce quasi-periodic patterns in the detected emissions. We obtain that among $21, 070$ simulated flares, $813$ events endure over 50 seconds, scaled with the temporal resolution of the Yohkoh Hard X-ray Telescope, and about $70 \%$ of these rather long-lasting events exhibit QPPs. We also illustrate that the applied CA model provides a wide range of periodicities for QPPs. Furthermore, we observe the presence of multiple periods in nearly $50 \%$ of the cases applying the Lomb-Scargle periodogram. A lognormal distribution is fitted to the unimodal distribution of the periods as a manifestation of an underlying multiplicative mechanism that typifies the effect of the system's independent varying parameters. The global maximum of the periods' lognormal distribution is located at $29.29 \pm 0.67$ seconds. We compare statistics of the simulated QPPs with parameters of the host flares and discuss the impacts of flare properties on QPPs’ periods. Considering the intrinsic characteristic of CA models, namely the repetitive load/unload mechanism, and the obtained pieces of evidence, we suggest that CA models may generate QPPs. We also examine the applicability of the autoregressive integrated moving average models to describe the simulated and observational QPPs.
\end{abstract}

\section{Introduction}\label{sec:intro}
Quasi-periodic pulsations (QPPs) with periods ranging from a few seconds to several minutes are often observed across the entire flare emissions, synchronously \citep{dennis1985, nakariakov2009, van2016quasi, dominique2018, nakariakov2018, hayes2019, zimovets2021}. The understanding of QPPs and their underlying physical mechanism has been subject to many studies in recent decades \cite[see][for a recent review]{mclaughlin2018}. In general, QPPs are believed to be generated by either propagation and dissipation of MHD modes in the solar corona (i.e., oscillatory processes) or an underlying load/unload mechanism (i.e., self-oscillatory processes). Each mechanism involves individual characteristics for these modulated patterns.

The main idea of the oscillatory processes is that small perturbations in plasma parameters (e.g., magnetic field strength, temperature, etc.) produce magnetohydrodynamic (MHD) waves. The evanescence of MHD waves modulates the plasma emissions, namely QPPs occur. Various models have been developed to investigate different oscillatory modes as possible candidates for energy modulations \cite[e.g.,][]{nakariakov2004b, nakariakov2005, chen2006, nakariakov2006, nakariakov2009, de2012, takasao2016, dennis2017}. According to this perspective, all MHD modes can modulate microwave emissions. The longitudinal and torsional modes could not describe the simultaneous modulation of microwave and HXR bands. Nevertheless, sausage and kink modes could produce synchronized modulation across the entire electromagnetic spectrum. These models are capable of producing a wide range of periodicities as well as describing multiple periods observed in QPPs \citep{nakariakov2003, melnikov2005, van2011, hong2021, lu2021}.

In self-oscillatory processes, the system is governed by a cyclical mechanism in which energy accumulates as a power supply (e.g., photospheric motions) slowly and continuously derives the system. If the free energy exceeds some threshold, the system releases a significant amount of energy through an avalanche process. In this context, the QPPs are triggered as a result of the frequent occurrence of magnetic reconnections which modulates the released energy and the acceleration of charged particles. The load/unload mechanism could describe synchronous emissions at different wavelengths \citep{aschwanden1987, craig1991, ofman2006p, zaitsev2008, mclaughlin2012p, mclaughlin2012, hayes2016, li2020quasi}.

Despite many theoretical and observational studies, it is yet unclear whether QPPs are derived by restoring forces in the perturbed coronal plasma, or they are direct outputs of repetitive reconnection regimes. It is also possible that each of these mechanisms effectively plays a role. Statistical analysis of oscillatory parameters and their governing scaling laws could practically provide information about the formation of QPPs. In addition to theoretical modeling and data analysis, numerical studies may provide a different and new approach to interpret QPPs.

Taking the idea of self-oscillatory processes into account, we perform a numerical study to simulate and investigate the statistics of QPPs. For this purpose, we consider solar flares as a self-organized critical (SOC) system and apply a modified version of the \cite{LH1991} model to numerically reproduce magnetic relaxations (flare-like events) and their accompanying QPPs. The remainder of this paper is organized as follows: In Section \ref{sec:load}, we argue the possibility of simulating QPPs as a result of a load/unload mechanism by cellular automaton (CA) models. Then, we introduce the characteristics of the applied model in Section \ref{sec:NSIM}. We describe the detection of QPPs in the numerical modeling in Section \ref{sec:stst}. We also present the statistics of simulated QPPs and compare them with observational reports. In Section \ref{sec:tsmodel}, we briefly review some classes of linear processes applicable in time series modeling and investigate their productivity in examining both simulated and observational QPPs. Finally, we conclude in Section \ref{sec:con}.

\section{Is It Technically Possible to Simulate QPPs by CA Models?}\label{sec:load}
By definition, the natural tendency of a system to automatically adjust its components to establish a critical state is called SOC, provided that exceeding some threshold leads to scale-free fluctuations in the system \citep{aschwbook2011}. In the SOC mechanism, a gradual energy supply drives the system towards a critical state at which it relaxes through a sequence (avalanche) of nonlinear energy-dissipative events. The frequency-size distribution of released energies manifests a power-law-like behavior. The threshold is an intrinsic feature of SOC and the driving timescale is much larger than the avalanche timescale.

SOC was introduced by \cite{bak1987} who established the first CA avalanche model, namely the sand-pile model. The model employs a grid with nodal values representing the sand distribution in the system. The initial state is constructed using random numbers and the energy supply mechanism operates as grains gradually drop into randomly selected sites, which builds up local piles. In the case of exceeding a critical slope in the system, the sand grains fall off the unstable pile to reduce the slant and relax the system. Since the driving operation time is negligible compared to the avalanche timescale, in numerical modeling the driving mechanism halts once an instability occurs and automatically reactivates after the release of the free energy. Hence, the energy balance is maintained in the SOC systems. The CA approach provides the ability to investigate the complex behavior of avalanche processes by breaking them down into smaller pieces.

The scale-free distribution of solar flare energies evokes the application of SOC models to study the stochastic nature of these events \citep{georg1998, mike2001, charbon2001, Litvinenko2001, nita2002, Mike2008, morales2008a, morales2010, mendoza2014, dani2015, farhang2018, alipour2019, farhang2020}. Solar flares occur due to a gradual and continual increase of magnetic stress in the solar atmosphere. The coronal magnetic field evolves as magnetic structures emerge, twist, braid, or annihilate on the Sun's surface, and relaxes through e.g., magnetic reconnections. Local relaxations may raise the stress elsewhere in the system and lead to a sequence of reconnections with massive release of energy \citep{Gold1960, forbes1991, longcope1996, biskamp2000, priest2000, somov2010, Loureiro2016}. As a result of the energy load and unload cycle, coronal magnetic topology changes. The stored magnetic energy converts to heat (e.g., ohmic dissipation) and kinetic energy of charged particles (generating MHD oscillatory modes, shock waves, and turbulence in the plasma). Therefore, various physical processes trigger thermal (EUV and SXR) and non-thermal (e.g., microwave, HXR, etc.) emissions during a solar flare \citep{Carmichael1964, sturrock1968, hirayama1974, kopp1976, somov1991, svestka1992, somov1997, fletcher2011, priest2014, Benz2017}. These emissions are often associate with QPPs.

\cite{Parker1988} proposed that gradual perturbations in the solar atmosphere lead the coronal magnetic structures towards an unstable state. Hence, small bursts (magnetic reconnections) occur as the building blocks of flaring events. Accordingly, \cite{LH1991} introduced a lattice-based model to investigate the efficiency of the CA approach in simulating solar coronal explosive events. Lu and Hamilton defined a discrete magnetic field over a 3D grid of nodes and applied a driving mechanism for topological evolution of the solar magnetic field. In the constructed model, the stressed magnetic field relaxes along with a series of magnetic reconnections, after which the system dives into a new equilibrium state. The frequency-size distribution of simulated flaring events is shown to follow a power law.

Based on the above discussion, CA models are technically adequate for simulating QPPs by a load/unload mechanism since they are well capable of reproducing repetitive reconnection regimes. Here, we use a modified version of the \citeauthor{LH1991} model in two dimensions and reassess its outputs and their physical interpretations. In our simulation, instead of a magnetic field we define a magnetic vector potential field over a 2D grid. Therefore, the magnetic field remains divergence-free over time. The model is introduced in detail in the next section.

\section{Model Properties}\label{sec:NSIM}
We consider a 2D lattice as a cross-section of a coronal loop and study the variation of the magnetic field inside this sector. Uniformly distributed random numbers are used to construct the initial state. The nodal values represent the magnetic vector potential field, $ \textbf{A}=A(x,y)\hat{z} $ in which $\hat{z}$ is a unit vector along the loop axis. We also consider the open boundary condition by keeping $ A = 0 $ on borders. This allows the energy to leak out of the system's boundaries. Otherwise, there will be an exponential growth in the energy and the system could never reach a stable state.

To imitate the evolution of the coronal magnetic field, a local driving mechanism is applied. Therefore, a small disturbance, generated from a uniform distribution, is added to an arbitrary node at each \textit{driving step}. Then, the stability of the entire system is checked through the criterion:
\begin{equation}
\label{eqb1}
\left| \Delta A_{i,j}\right| \hspace{1mm}\equiv \hspace{1mm} \left| A_{i,j} - \frac{1}{4}\sum_{l=1}^{4} A_{l}\right| > A_{c},
\end{equation}
where the sum runs over the four nearest neighbors $ (A_{l}) $. The instability threshold, $ A_{c}, $ is a random number generated from a Gaussian distribution with e.g., $\mu=1$ and $ \sigma=0.01 $. Should the instability criterion fulfil anywhere in the system, the field locally relaxes through a set of redistributions:
\begin{eqnarray}
\label{eqb2}
A_{i,j}^{n+1}=&&A_{i,j}^{n}- \frac{4}{5} A_{c}, \vspace{13mm} \nonumber \\
A_{l}^{n+1}=&&A_{l}^{n}+ \frac{1}{5}A_{c},
\end{eqnarray}
where $ n $ denotes the \textit{evolution step}, $A_{i,j}$ is the central node, and $ l=1,2,3,4 $.

In this perspective, a magnetic reconnection is regarded as short-range interactions between an unstable node and its nearest neighbors. A succession of redistributions from an inceptive instability somewhere to ultimate stability everywhere is called an avalanche (flare). Various choices of redistribution rules are possible including conservative or nonconservative, isotropic or anisotropic, and deterministic or probabilistic. Also, long-distance interactions are conceivable \cite[see:][]{isliker1998, charbon2001, stru2014, farhang2018, farhang2019}.

There are two different aspects in measuring the constructed configuration's energy: the lattice energy and the released energy. The development of the former should expectedly demonstrate an energy balance in the system as the implemented open boundary condition avoids a continuous increase of energy. Therefore, the system reaches an approximate stationary state on top of which the energy fluctuates due to excitements and relaxations. Besides exhibiting a power-law-like behavior, the latter could provide important information about the underlying physical mechanisms.

In a magnetic configuration, the energy is proportional to the square of the magnetic field strength. Therefore, we have:
\begin{eqnarray}
\label{eqbb3}
E_{\rm latt} \propto \sum_{\textrm{lattice}} B^2 \propto \sum_{\textrm{lattice}} |\nabla \times \textbf{A}|^{2} \propto \sum_{\textrm{lattice}} A^{2}.
\end{eqnarray}
The matter of interest is to appraise the ability of CA models in generating QPPs. An extensive understanding of the energy release process is required to achieve this purpose. The amount of released energy during each redistribution is:
\begin{eqnarray}
\label{eqb3}
{e}_{i,j}=\sum {\left({A}^{n+1}\right)}^{2}-\sum {\left({A}^{n}\right)}^{2}= \frac{4}{5}\left(1 - 2 \frac{| {\rm\Delta }{A}_{i,j}|}{{A}_{c}}\right){A}_{c}^{2},
\end{eqnarray}
where the sum runs over the five nodes engaged in the redistribution.

Since there is no preferred direction in the system, all unstable nodes are redistributed simultaneously. One can consider it as a bunch of reconnecting events occurring in a single \textit{time frame}. Therefore, the total released energy during each frame is:
\begin{eqnarray}
\label{eqb4}
e_{frame} = \sum e_{i,j},
\end{eqnarray}
and the sum includes all the identified unstable nodes.

Local redistributions may cause other instabilities in the system. Thus, the system's stability needs to be examined again after each time frame. In the case of appearing new instabilities in the system, the whole \textquotedblleft check-redistribute\textquotedblright~procedure continues until an equilibrium is achieved. Then, the driving mechanism operates again. Accordingly, the flare energy is the sum of the released energies in successive frames between two driving steps or equivalently the energy difference between two consecutive driving steps:
\begin{eqnarray}
\label{eqb5}
E_{\textrm{flare}} = \sum e_{frame} \simeq E_{\textrm{latt}}^{t+1} -E_{\textrm{latt}}^{t},
\end{eqnarray}
where $ t $ denotes the driving step.

\section{Numerical QPPs and their Periodicities}\label{sec:stst}
\begin{figure}
\includegraphics[width=18cm,height=9.5cm]{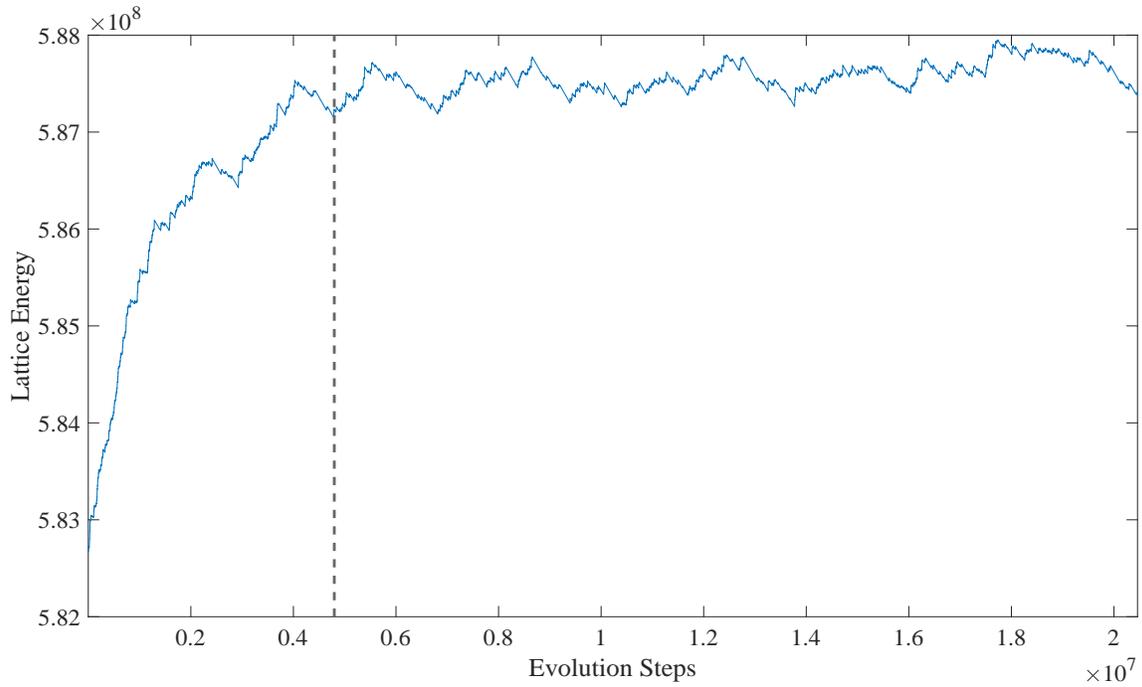}
\caption{The evolution of the lattice energy in a magnetic vector potential lattice (with $2^{16}$ grids) during 200,000 driving steps. The initial lattice is constructed using uniformly distributed random numbers. After nearly 5 million evolution steps (dashed-line) the system reaches to an approximate stationary state. During the following steps, the driving mechanism injects energy into the system. Whereas, the avalanches emit energy.}
\label{Fig1}
\end{figure}

We established a square lattice of magnetic vector potential field with $\bm{256 \times 256}$ grids and studied its evolution over $200,000$ driving steps. Figure \ref{Fig1} shows the variation of the lattice energy during which $20$ million, $275$ thousand, $220$ magnetic reconnections occurred in the system. Starting from an initially random state, it took over $4$ million evolution steps for the system to reach an approximate stationary state over which the energy fluctuates around a constant average value. We restrict our study to the total number of $21,070$ flaring events registered after the energy balance is delivered to the system (i.e., $124,064$ driving steps or equivalently $15,677,989$ evolution steps).

\subsection{QPPs in CA Models}\label{sec:consept}
\begin{figure}
\includegraphics[width=18cm,height=9.5cm]{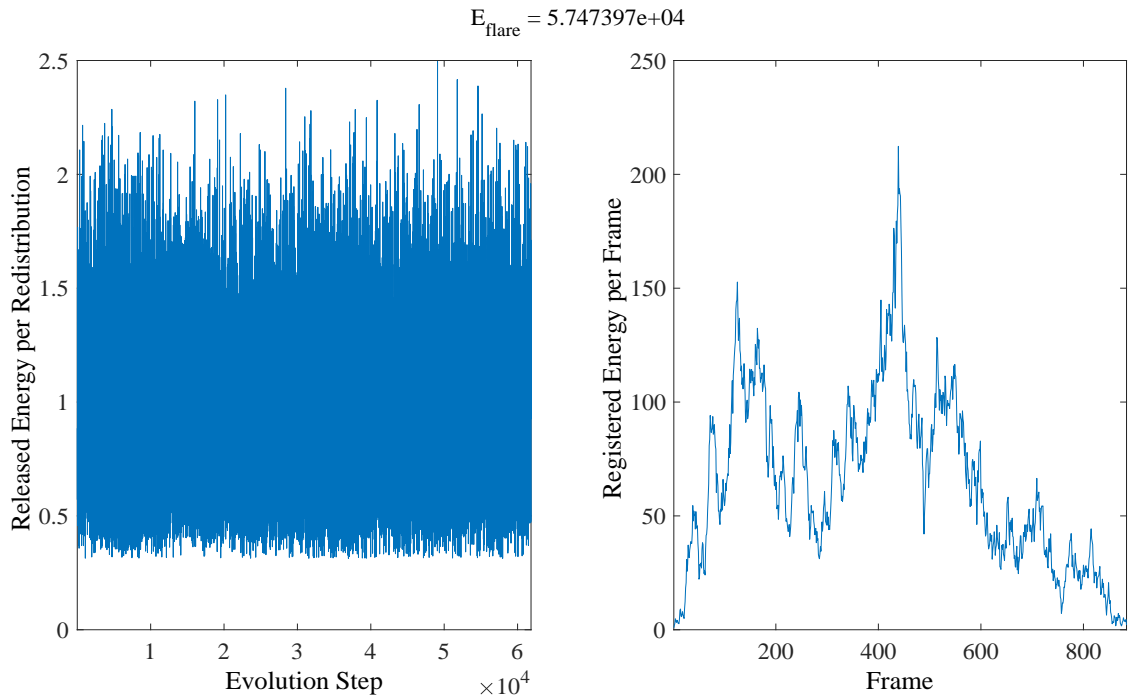}
\caption{The registered energies of a typical simulated flare. During the subject flare, $61,834$ magnetic reconnections occurred (within 884 frames) leading to a total energy release of $ 5.74 \times 10^{4} $. The left panel shows the amount of energy released at each magnetic reconnection. The right panel illustrates the detected energies in successive frames.}
\label{Fig2}
\end{figure}

We propose that the released energies during individual frames can be considered as the energy recorded by solar detectors within their temporal resolution. Hence, the released/detected energy from a flaring event is supposed to exhibit a quasi-periodic pattern (as a representation of QPPs). Figure \ref{Fig2} displays an example of the detected QPPs for a simulated flare with the total dimensionless energy of $ 5.74 \times 10^{4} $, in which $61,834$ magnetic reconnections occurred during 884 frames.

For comparison, the observed emissions of a solar flare recorded by the Yohkoh satellite are presented in Figure \ref{Fig3}. The subject flare occurred on March 1th, 1998 at 17:09 and lasted over $10$ minutes. The Yohkoh Hard X-ray Telescope (HXT) registered this event in four spectral bands (i.e., L, M1, M2, and H bands) with a temporal resolution of half-seconds. The attended oscillatory patterns are referred to as QPPs.

One important question is whether all registered energies within consecutive frames in both observations and simulations exhibit quasi-periodic characteristics. \cite{inglis2016} conducted an extensive study on several M and X flares recorded by the Geostationary Operational Environmental Satellite (GOES) and the Gamma-ray Burst Monitor (GBM). They found that roughly one-third of the studied flares were attended by quasi-oscillatory patterns. Furthermore, \cite{szaforz2019} investigated the partially occulted flares listed in the Yohkoh Legacy Data Archive (YLA) and determined that less than $ 50\% $ of these emissions exhibit QPPs. Failure to detect QPPs in a large fraction of observations might be due to limited sensitivity of the instruments, deficiency of computational methods, inappropriate interpretation of QPPs, or the absence of these pulsations along with flaring events. In the following, we assess the likelihood of numerical QPPs representing similar statistics. But first, we introduce the algorithm used to study the periodicities.

\subsection{The Lomb-Scargle Periodogram}\label{sec:plomb}
The Lomb-Scargle periodogram is a powerful mathematical tool developed to study the oscillatory parameters in data samples. Having its origins in the Fourier transform, this periodogram decomposes a discrete regular/irregular signal ($ S $) to its independent frequency components:
\begingroup
\Large
\begin{eqnarray}
\label{eq7}
\mathcal{P}_{S} = \frac{1}{2}\bigg(\tfrac{{[\sum_{j}{S}_{j}\cos\omega({t}_{j}-\tau)]}^{2}}{\sum_{j}{\cos}^{2}\omega({t}_{j}-\tau)}+ \tfrac{{[\sum_{j}{S}_{j}\sin\omega({t}_{j}-\tau)]}^{2}}{\sum_{j}{\sin}^{2}\omega({t}_{j}-\tau)}\bigg),
\end{eqnarray}
\endgroup
where the sum includes all data samples and $ \mathcal{P}_{S} $ is the estimate of the power spectral density (PSD). The parameter $ \tau $,
\begin{eqnarray}
\label{eq8}
\tan(2\omega\tau) = \frac{\sum_{j}\sin 2\omega{t}_{j}}{\sum_{j}\cos 2\omega{t}_{j}},
\end{eqnarray}
guarantees the orthogonality of the sinusoidal terms in Equation (\ref{eq8}) and equips the algorithm to be invariant under global time shifts, in contrast to the fast Fourier transformation \citep{lomb1976,scargle1982}.

By some means, the Lomb-Scargle periodogram is equivalent to other frequency analysis techniques, namely the least-square fitting, phase-folding, and Bayesian methods \citep{vanderplas2018}. It also provides a $ \chi^{2}- $distributed power spectrum for Gaussian uncertainties \citep{vio2013}. Due to its convenient properties, the Lomb-Scargle periodogram has a widespread application in time series analysis particularly in astronomical studies \cite[e.g.,][]{tarnopolski2021, zhang2021, saikia2022}. Here, we apply this technique to study the numerical QPPs' periodicities.

Once the analysis is performed, the significance level of each frequency component is validated against the \textquotedblleft false alarm probability\textquotedblright~(FAP). The FAP estimates the likelihood of a frequency to appear in the periodogram due to the existence of noise in the signal rather than the original source. Various methods have been developed to measure the FAP in presence of different contaminative sources \citep{horne1986, baluev2008, delisle2020, delisle2020_2}. In the simplest case, the FAP is measured assuming a white noise in the time series, whilst in the more complicated contexts correlated noise are considered.

Figure \ref{Fig4} shows an example of a PSD obtained for the observational QPPs together with the FAP significance levels of $50, 10, 1,$ and $0.01$ percent. To identify the dominant frequencies in the power spectrum, we use a Gaussian filter:
\begin{eqnarray}
\label{eq9}
F(\mathcal{P},\nu)=H(\nu)\exp\left(- \frac{{\mathcal{P}-C(\nu)}^{2}}{{W(\nu)}^{2}} \right),
\end{eqnarray}
where $ H, W, $ and $ C $ are the height, width, and contribution of each peak in the PSD. We also filter out the harmonics. Further details of the applied frequency extraction technique are discussed in the next section.

\subsection{Statistics of the Simulated flares and QPPs}\label{sec:res}
Applying the CA approach, $21,070$ flares are simulated with dimensionless energies of $0.7$ to $2.4\times 10^{5}$. Figure \ref{Fig5} displays the probability distribution functions (PDFs) of the simulated flare energies and lifetimes (durations, $D$). Expectedly, both distributions exhibit power-law-like behaviors \citep{LH1991, charbon2001}. The simulated events are categorized into $5$ types (labeled as A, B, C, M, and X flares) relevant to the GOES classification system. Among all events, only $813$ flares (about $3.8\%$) released their energy in more than 100 frames. Therefore, considering the liberated (free) energies in consecutive frames analogous to emissions recorded by the HXT in half-seconds intervals, most of the flares lasted less than $50$ seconds. The simulated events with $D>50$ seconds are mostly C-class and above. Durations less than $50$ seconds may also relate to small-scale flaring events such as small-scale brightenings (microflares, campfires, bright coronal points, etc.) that have been recently observed with space missions \cite[see e.g.,][]{chen2021, berghmans2021, shokri2022}.

As shown in Figure \ref{Fig5}, both distributions are fitted with power-law functions. For the energy distribution the power index is $ 1.45 \pm 0.02 $ and the goodness-of-fit is assessed using the Kolmogorov-Smirnov test (KS-test). The KS-test gives a measure of the departure between two distributions based on a decision-making routine (between a null and an alternative hypothesis). It also returns a $p-$value that validates the decision. In this case, the null hypothesis is that the PDF follows a power law and the obtained $p-$value, $0.98$, does not reject the null. This power-law (scale-free) behavior suggests that the CA self-oscillatory processes can be considered as the generative mechanism of self-similar, self-organized or self-organized criticality. It also implies that magnetic reconnections are most probably the leading generator of flaring events \citep{Parker1988, aschwbook2011}. For the duration distribution, the power-law function exhibits a better match with the significant events (the tail of the distribution) rather than the small-scale events with energies $<$ B0-class, including A-class flares and other tiny features. The power index is $1.71 \pm 0.02$ with $p-$value of $0.90$. The obtained $p-$value indicates that the difference between the PDF and the power-law function is not statistically significant.

To perform a cyclical analysis, usually, the first step is to normalize the subject time series. Various forms of normalization are possible \citep{ogasawara2010, panigrahi2013}. Here, we apply two common types and argue their effects on the PSDs. The applied normalization routines are:
\begin{eqnarray}
&&S_{N} = \frac{S-\bar{S}}{\bar{S}}, \label{eq10}\\
&&S_{MA} = \frac{S-\hat{S}_{k}}{\hat{S}_{k}}\label{eq11},
\end{eqnarray}
where $\bar{S}$ is the overall mean value of $S$ and $\hat{S}_{k}$ is the moving average over a window with size $k$. In the remaining of this paper, $S_{N}$ and $S_{MA}$ are referred to as normalized and smoothed time series, respectively.

Figure \ref{Fig4} displays the PSDs obtained for the HXR emissions of a solar flare recorded by Yohkoh together with the FAP levels. The FAP provides an estimation of the accuracy of the extracted periods. Hence, we only consider periods with intensities higher than $ 0.01 \%$ significance level (the green horizontal line) as true findings. The top panels of the figure are the results of applying the Lomb-Scargle periodogram on the smoothed times series with $k=80$. The most prominent peaks of the PSDs are located at $22.24, 21.83,$ and $27.48$ seconds for L, M1, and M2 bands, respectively. No significant peak is obtained for the H band. The bottom panels correspond to the normalized time series. In contrast with the smoothed time series that most likely have one prominent peak, the PSDs of normalized light curves might manifest more peaks. However, the extracted periods of the smoothed samples are highly sensitive to the averaging window size and various choices of $k$ do not reproduce convergent results. Therefore, Equation (\ref{eq10}) provides a more reliable normalization (without any information loss), particularly in cyclical analysis. We apply Equation (\ref{eq10}) to the registered energies and evaluate their periodicities. We include all the peaks above the maximum FAP level in the performed survey.

Figure \ref{Figm} illustrates an example of the obtained PSDs for a few of simulated flares. As seen in the figure, the power spectrums exhibit a power-law-like behavior in the log-frequency presentation. Such a statistic of the PSD have been previously reported in the solar and stellar studies including observations of solar flares and QPPs \cite[e.g.,][]{cenko2010,gruber2011,ireland2014,inglis2016}. This result also attests to the productivity of the CA approach in the reproduction of flares and QPPs.

We investigate the properties of the simulated QPPs in two regimes of $D\leq 50$ and $D>50$ seconds. Figure \ref{Fig6} displays the histograms (number) of extracted periods for both regimes individually (panels a \& b) and collectively (panel c). Among $20,257$ events of the first regime, about $36\%$ of flares lasted long enough (more than 5 frames) to perform the frequency analysis. Our analysis shows that, less than $1\%$ of these flares are accompanied by QPPs. However, in the seconds regime ($D>50$), quasi-periodic patterns are observed in nearly $70\%$ of events. We observe that about $47\%$, $41\%$, and $12\%$ of QPPs exhibit $1, 2,$ and $3$ conspicuous peaks in their PSDs, respectively. We fit a lognormal distribution to the unimodal histograms of Figure \ref{Fig6}. The lognormal distribution is,
\begin{eqnarray}
L(p,\mu,\sigma) = \frac{1}{p\sigma\sqrt{2\pi}}\exp\bigg( \frac{-(\ln(p)-\mu)^{2}}{2 \sigma^{2}} \bigg), \label{eq12}
\end{eqnarray}
where $\mu$ and $\sigma$ are the scale and shape parameters, respectively. The global maximum of the fitted lognormal distribution is:
\begin{eqnarray}
Mode[p]=e^{\mu-\sigma^{2}},
\label{eq14}
\end{eqnarray}
which represents the most probable period of the simulated QPPs and equals to $33.66 \pm 0.71$ and $29.29 \pm 0.67$ seconds for panels (b) and (c), respectively. Applying the KS-test, the $p-$values are also obtained $0.56$ and $0.62$, respectively.

It is well known that, the lognormal distribution is a characteristic of stochastic systems that associate with the multiplicative effect of independent varying parameters \citep{mitzenmacher2004, tokovinin2014, ruocco2017}. Therefore, the obtained lognormal distribution for the periodicities might originate in the driving mechanism, the number of unstable sites contributing to reconnections, or even the amount of released energies. One may ask how two different generative mechanisms i.e., the power-law behavior (Figures 5 and 6)  and the lognormal behavior (Figure 7) may work together in a system. \cite{pauluhn2007} developed a stochastic model for generating small-scale flaring emissions with an initial power-law distribution that evolved via random kicks as a multiplicative process. In such a system, although the time series develops due to a multiplying generative mechanism, the inceptive power-law behavior of the event distribution is stored in the memory of the time series. The machine learning methods have been applied to determine the power-law index of such simulated and observational time series \cite[e.g.,][]{bazargan2008,taj2012,sadeghi2019,upendran2021}.

\cite{valdivia2005} argued that intermittent SOC phenomena (e.g., reconnections and substorms) and self-similar turbulent plasma sheets present power-law characteristics. Furthermore, \citeauthor{alipour2022} studied the morphological and intensity structures of the small-scale brightenings (campfires) observed by the Solar Orbiter. They discussed that, to some extent, both the power law and lognormal functions could conveniently model the heavy-tailed distributions of campfires' duration, peak intensity, and size. Moreover, they showed that small-scale brightenings mainly occur at the supergranular cell boundaries (Figure 7, therein). Therefore, these features are analogous to flaring events in the sense of an underlying generative mechanism (i.e., magnetic reconnections), except that they appear in much smaller scales. In other words, one may consider QPPs, or even campfires, related to the reconnection regimes that occur between adjacent magnetic structures as the magnetic field sweeps towards the footpoints via horizontal or turbulent flows. For further discussion on power laws and multiplicative processes see \cite{stefanthurner2018}, section 3.3.3 therein.

Evaluating the existence of any statistical relationship between the oscillatory parameters of the QPPs and the host flares could provide intuition about the underlying generative mechanism. Here, we assess the dependency of the simulated QPPs' periods on the host flare durations and energies. Table \ref{table1} presents the correlations between these variables. The obtained positive covariances indicate the dependency of the QPPs' periods on the flare parameters. However, the Pearson coefficient does not confirm a strong linear correlation. Figure \ref{Fig7} presents the scatter plots of the extracted periods versus the flare durations (top panel) and energies (bottom panel). The fan-shaped diagram of the top panel partially corresponds to the repeated values as multiple periods are detected in the PSDs of some QPPs (see e.g., black rectangles in the figure). It may also indicate the existence of another covariate influencing the dependency of pulsation periods on the flare durations. Nevertheless, no particular pattern is observed in the bottom panel.

We also calculate the non-linear correlations between the QPPs' periods and flare properties using the mutual information \citep{shannon1948,kreer1957}:
\begin{eqnarray}
\mathcal{I}(p;f)=\int\int P(p,f)\log \frac{P(p,f)}{P(p)P(f)}dp df,
\label{eq15}
\end{eqnarray}
where, $P(p)$ and $P(f)$ are marginal distributions of the periodicities and flare properties, respectively and $P(p,f)$ is the joint distribution. Following the \cite{peng2005} routine, the mutual information is obtained a positive value (for both the duration and energy, see Table \ref{table1}) which implies the existence of a missing parameter responsible for the observed non-constant (heterogeneous) variances. Although the measured correlations attest to the impacts of flaring events (reconnection regimes) on the production of QPPs, further observational or theoretical investigations are required to fully understand the involved processes and determine the parameters affecting the QPPs characteristics.

\begin{table}
\caption{The statistical correlations between the QPPs' periods (p) and the host flare properties, i.e., duration (D) and energy (E).}
\begin{center}
\begin{tabular}{c c c c}
\hline
Variables & Covariance & Pearson Correlation & Mutual Information\\
\hline
\hline
(p,D) & $4.1$ $\times$ $10^{3}$ & $0.55$ &  $5.17$ \\
(p,E) & $6.8$ $\times$ $10^{5}$ & $0.63$ &  $6$ \\
\hline
\end{tabular}
\end{center}
\label{table1}
\end{table}

\section{Modeling The Time Series of QPPs}\label{sec:tsmodel}
The study of real-world data and synthetic times series has a rich history that goes back over a hundred years. Time series analysis provides statistical information that sheds light on the underlying generative mechanisms and may even result in prediction. To perform such analyses, numerous algorithms/tools have been developed which are mainly classified into two categories: frequency-based and time domain methods \citep{kantz2003, box2015, shumway, chatfield2019}. A well-established paradigm of the first, e.g., the Lomb-Scargle periodogram, is introduced and implemented in Section \ref{sec:stst}. The second type operates based on correlation analysis. Here, we study the stochastic time series of both simulated and observational QPPs using a parametric time domain method. We apply an autoregressive integrated moving average (ARIMA) model to the non-stationary time series of QPPs and investigate their efficiency in characterizing the QPPs behavior.

The ARIMA models could adequately study processes with time varying mean values. These models, first, remove the non-stationarity by applying a backshift operator and then, fit either an autoregressive (AR), a moving average (MA), or a combination of these terms on the subject series \citep[The Econometrics Toolbox User's Guide,][]{matlab}. Both the AR and MA terms cope with the serial autocorrelations of a given time series and use the past observations and errors of the sample, respectively, to make predictions for the future values. A MA process of order $q$ is defined as:
\begin{eqnarray}
S_{t}&=&Z_{t}+\theta_{1}Z_{t-1}+\dots+\theta_{q}Z_{t-q},\nonumber \\
&=&(1+\theta_{1}B+\theta_{2}B^{2}+\dots+\theta_{q}B^{q})Z_{t},
\label{eqapp1}
\end{eqnarray}
where $S_{t}$ is the sample value at time $t$, $\theta_{i}$ is the MA coefficient for $i=1..q$, $B$ is the backshift operator, and $Z_{t}$ is a random value, generated from a Gaussian distribution (white noise). For the simplicity and without loss of generality, the expectation of the noise is assumed to be zero. Similarly, an AR process of order $p$ is:
\begin{eqnarray}
S_{t}&=&Z_{t}+\phi_{1}S_{t-1}+\dots+\phi_{p}S_{t-p},\nonumber \\
&=&Z_{t}+(\phi_{1}B+\phi_{2}B^{2}+\dots+\phi_{p}B^{p})S_{t},
\label{eqapp2}
\end{eqnarray}
and $\phi_{i} ~ (i=1..p)$ is AR coefficient.

Considering Equations (\ref{eqapp1}) and (\ref{eqapp2}), an ARIMA$(p,d,q)$ process is defined as:
\begin{eqnarray}
\bigg( 1-\sum_{i=1}^{p} \phi_{i}B^{i} \bigg){(1-B)}^{d}S_{t}=\bigg( 1+\sum_{j=1}^{q} \theta_{j}B^{j} \bigg)Z_{t},
\label{eqapp3}
\end{eqnarray}
where $d$ is the difference parameter and takes integer values. ARIMA models indicate the existence of short-memory autocorrelations in a system. A more general approach is to consider an additional $1/f$-type long-memory autocorrelation in the time series, for which the autoregressive fractionally integrated moving average (FARIMA or ARFIMA) models are applicable. The FARIMA models share the same representation of Equation (\ref{eqapp3}) except that the difference parameter adopts non-integer values.

\begin{figure}
\includegraphics[width=18cm,height=11cm]{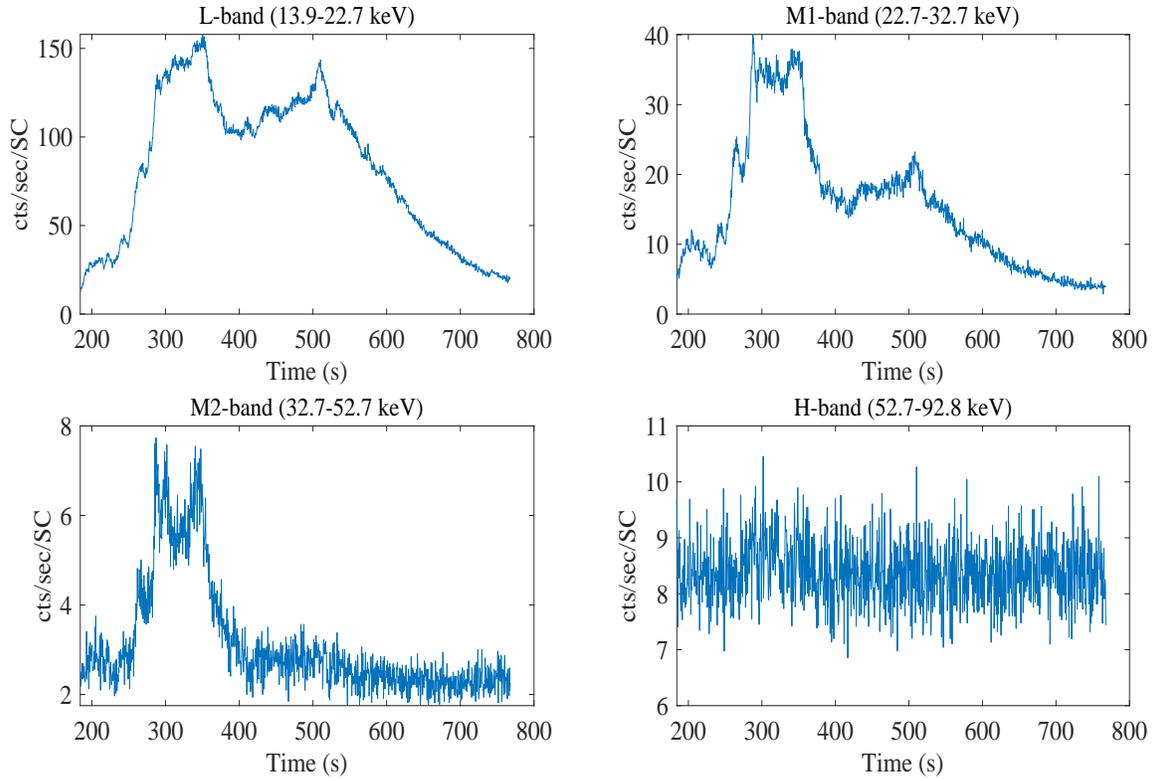}
\caption{The HXR emissions of a solar flare registered by the Yohkoh satellite. The flaring event occurred at 17:09 March 1th, 1998 and was recorded by the HXT instrument in different energy bands (L: 13.9\textit{-}22.7 keV, M1: 22.7\textit{-}32.7 keV, M2: 32.7\textit{-}52.7 keV, and H: 52.7\textit{-}92.8 keV). The vertical axis represents the average count rate over the 64 spatially-modulated subcollimators (SC).}
\label{Fig3}
\end{figure}

\begin{figure}
\centering
\includegraphics[width=18cm,height=11cm]{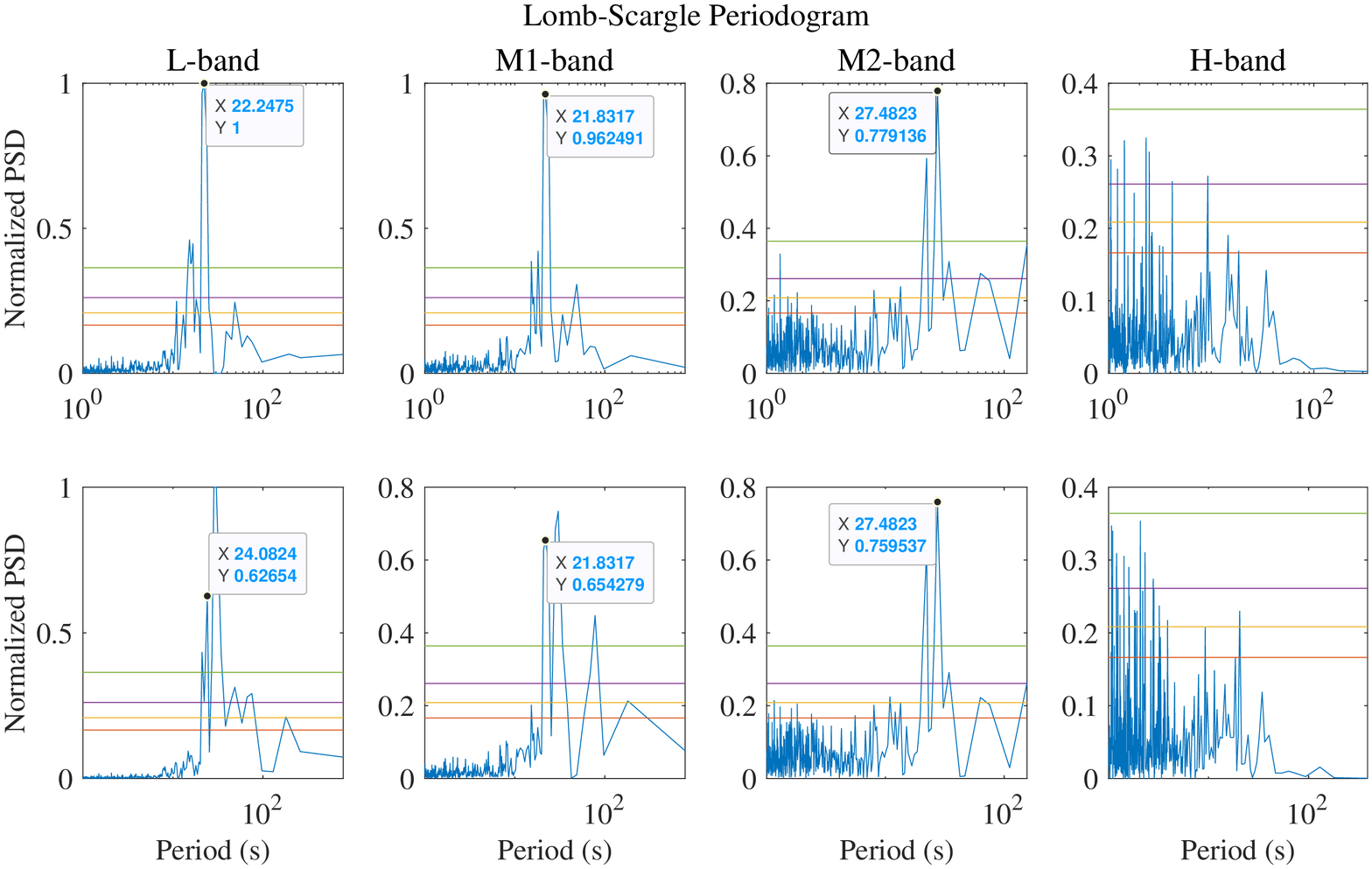}
\caption{The normalized PSDs of the HXR emissions of a solar flare (Figure \ref{Fig3}) registered in 4 different energy bands (L, M1, M2, and H) by the Yohkoh satellite. The frequency (period) components are extracted applying the Lomb-Scargle periodogram. The horizontal lines represent the FAP significance levels of $50, 10, 1,$ and $0.01$ percents displayed by the red, yellow, purple, and green colors, respectively. The top panels correspond to the smoothed time series (Equation \ref{eq11}) whilst the bottom panels are obtained for normalized time series (Equation \ref{eq10}).}
\label{Fig4}
\end{figure}

\begin{figure}
\centering
\includegraphics[width=8.5cm,height=9.5cm]{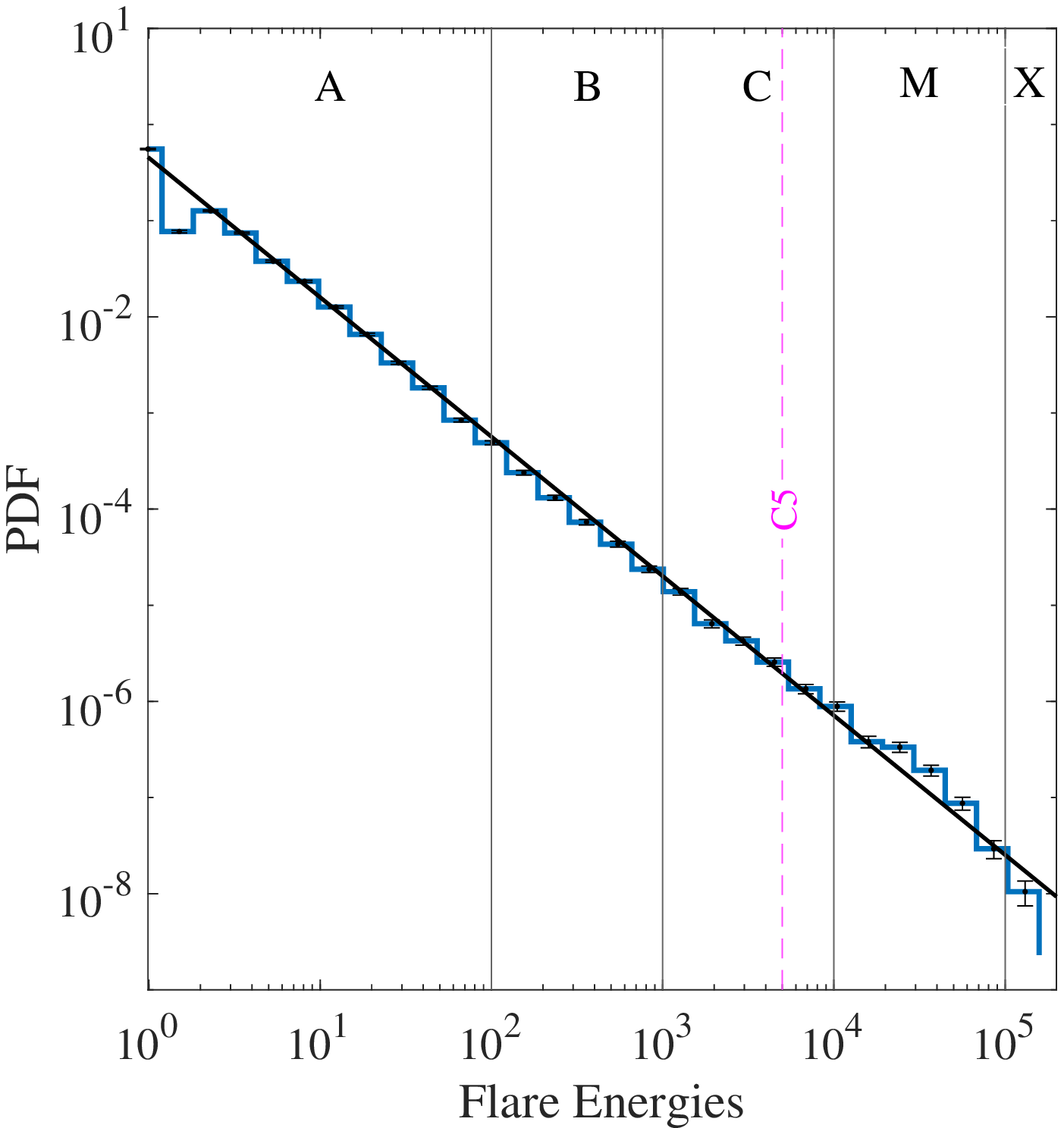}
\includegraphics[width=8.5cm,height=9.5cm]{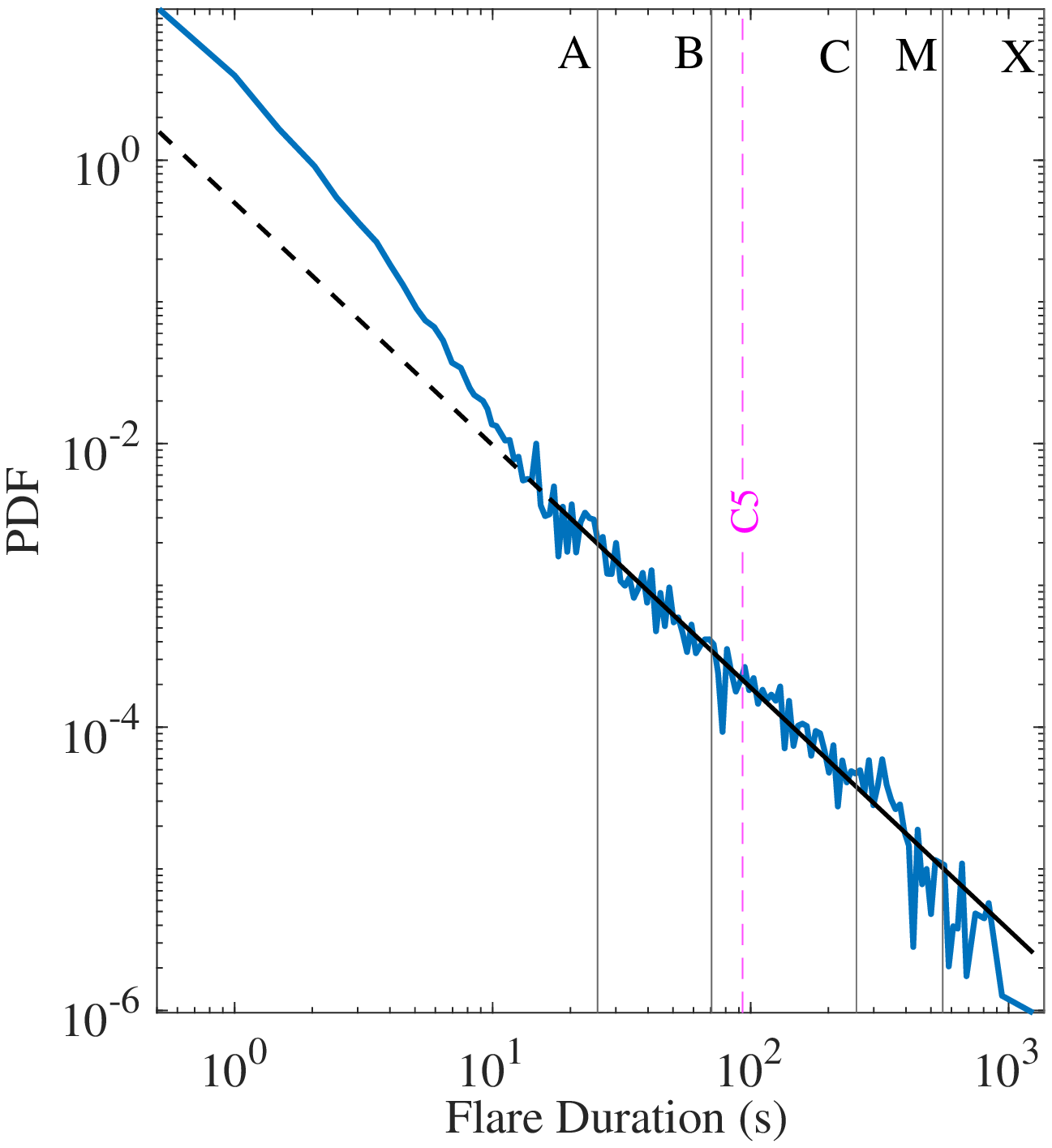}
\caption{The PDFs of the flare energies (left panel) and durations (right panel) of simulated flares. Types of flares (A, B, C, M, and X) are identified based on the GOES classification system. A power-law function is fitted to both distributions. The power indices are $1.45 \pm 0.02$ and $1.71 \pm 0.02$, with the $p-$values of $0.98$ and $0.90$ for the left and right panels, respectively.}
\label{Fig5}
\end{figure}

\begin{figure}
\centering
\includegraphics[width=18cm,height=6cm]{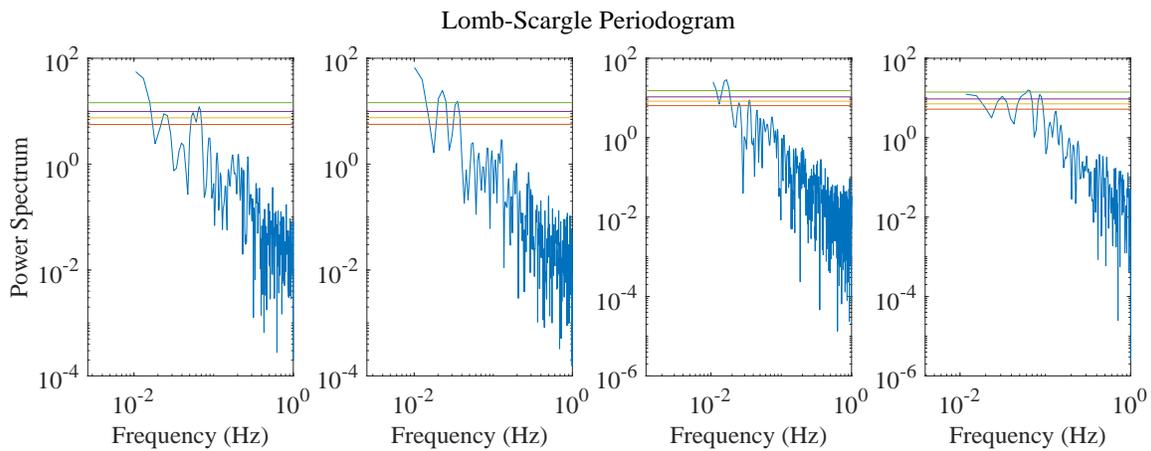}
\caption{The log-frequency presentation of the PSDs for some of the simulated events.}
\label{Figm}
\end{figure}

\begin{figure}
\centering
\includegraphics[width=18cm,height=11cm]{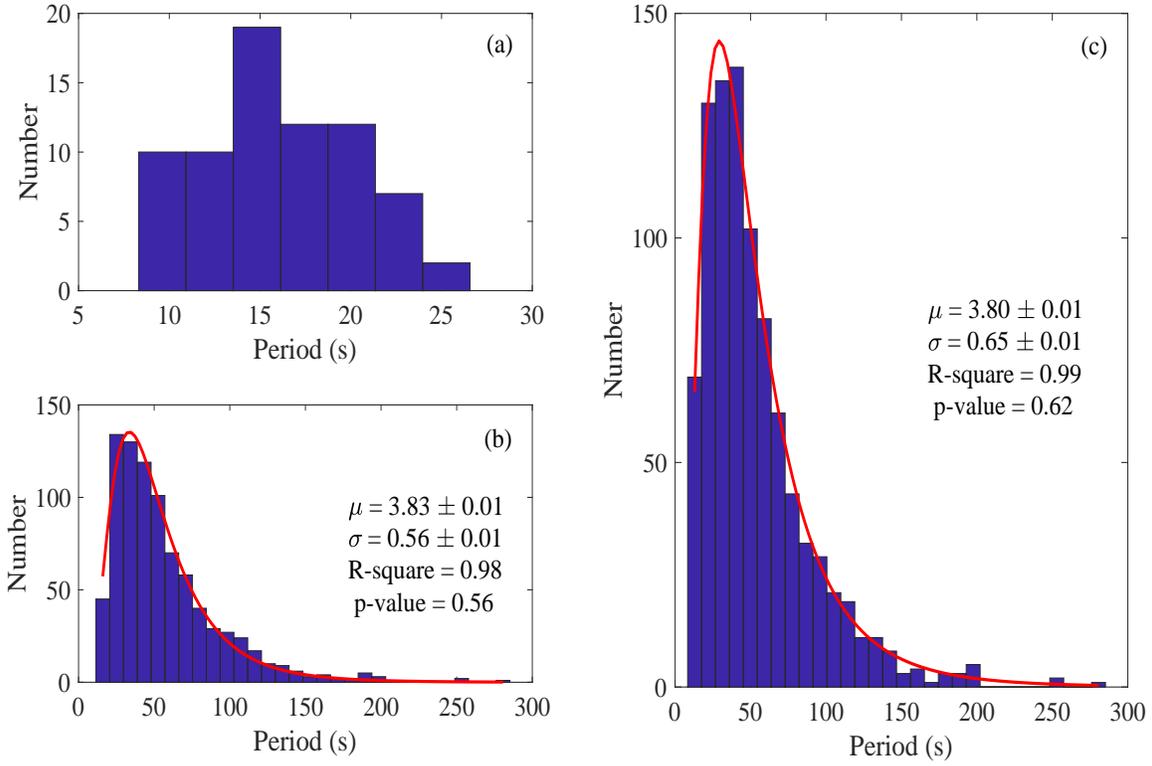}
\caption{The histograms of the extracted periods for the simulated QPPs for: (a) $D \le 50$ seconds and (b) $D > 50$ seconds. Panel (c) exhibits the number of collective periods of both regimes. The global maximums of the lognormal distributions are located at $33.66 \pm 0.71$ and $29.29 \pm 0.67$ seconds for panels (b) and (c), respectively. The goodness-of-fit is assessed applying both the R-squared statistics and KS-test.}
\label{Fig6}
\end{figure}

\begin{figure}
\centering
\includegraphics[width=18cm,height=10cm]{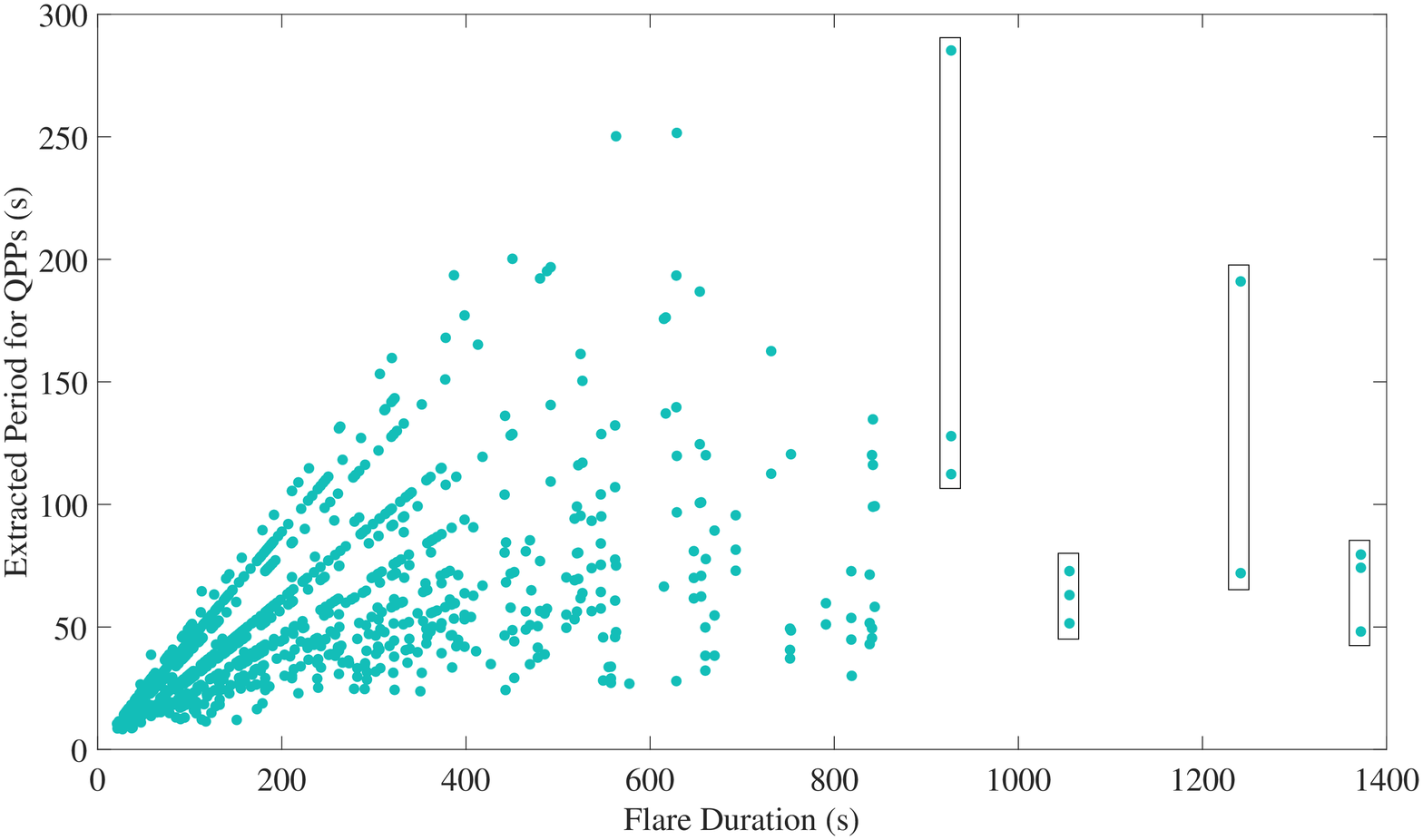}\\
\includegraphics[width=18cm,height=10cm]{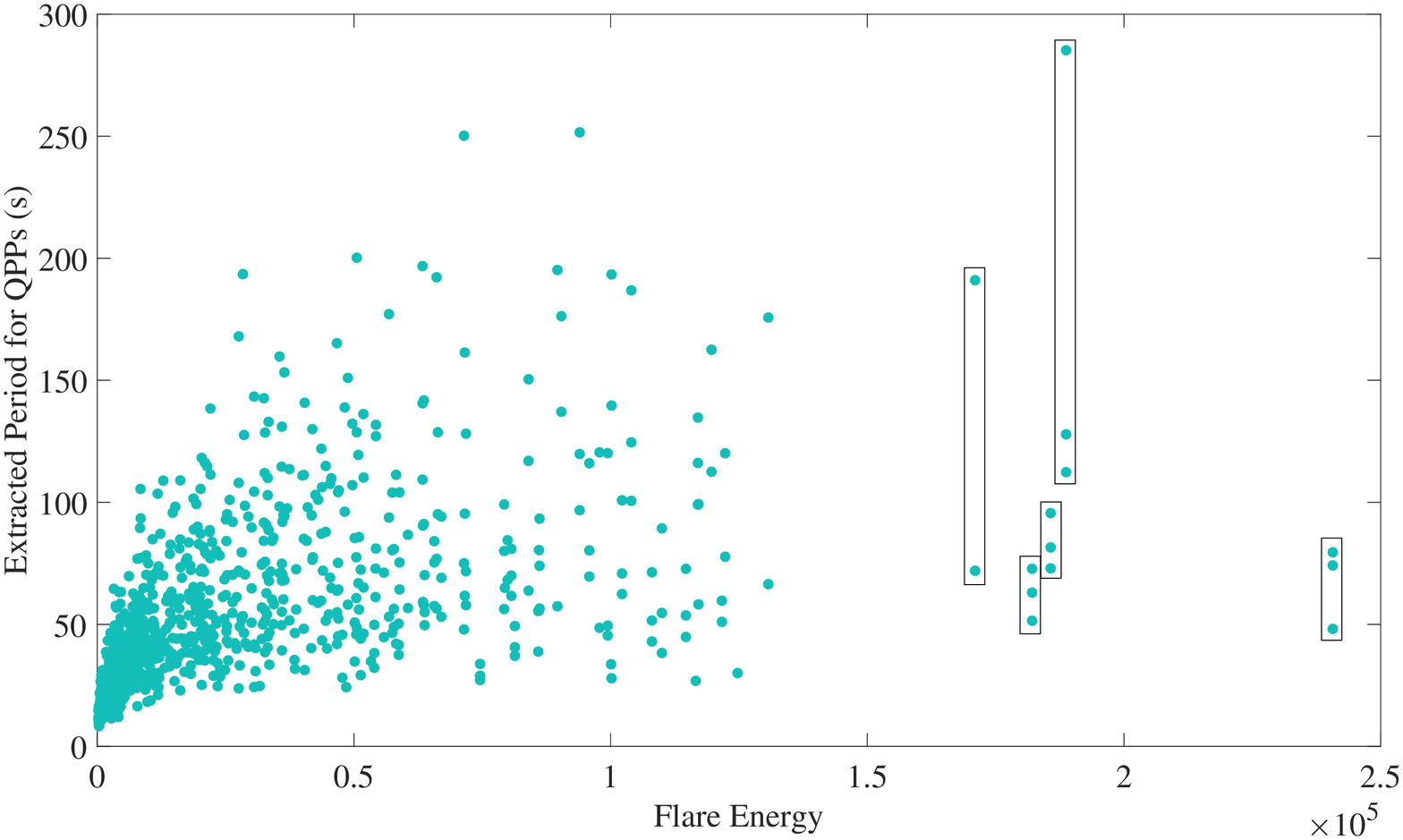}
\caption{The scatter plots of the extracted periods for the simulated QPPs versus the host flare durations (top panel) and energies (bottom panel). The black rectangles illustrate some examples of the QPPs for which multiple periods are detected.}
\label{Fig7}
\end{figure}

The application of the time domain methods has been of interest in astronomical studies in recent decades \cite[e.g.,][]{lazio2001dual,stanislavsky2009, templeton2009, feigelson2012modern, kelly2014, feigelson2018}. In the following, we perform a parametric analysis on the registered energies of both simulated and observational flares of Figures \ref{Fig2} and \ref{Fig3}, respectively. We use MATLAB's \textit{EconometricModeler} toolbox to examine several choices of $p, d,$ and $q$ for both ARIMA and FARIMA models and discuss the results. The main motivation for fitting an ARIMA/FARIMA model to the time series of QPPs is to investigate the functionality of these methods in describing the quasi-oscillatory patterns of solar flares.

\begin{figure}
\centering
\includegraphics[width=18cm,height=8.5cm]{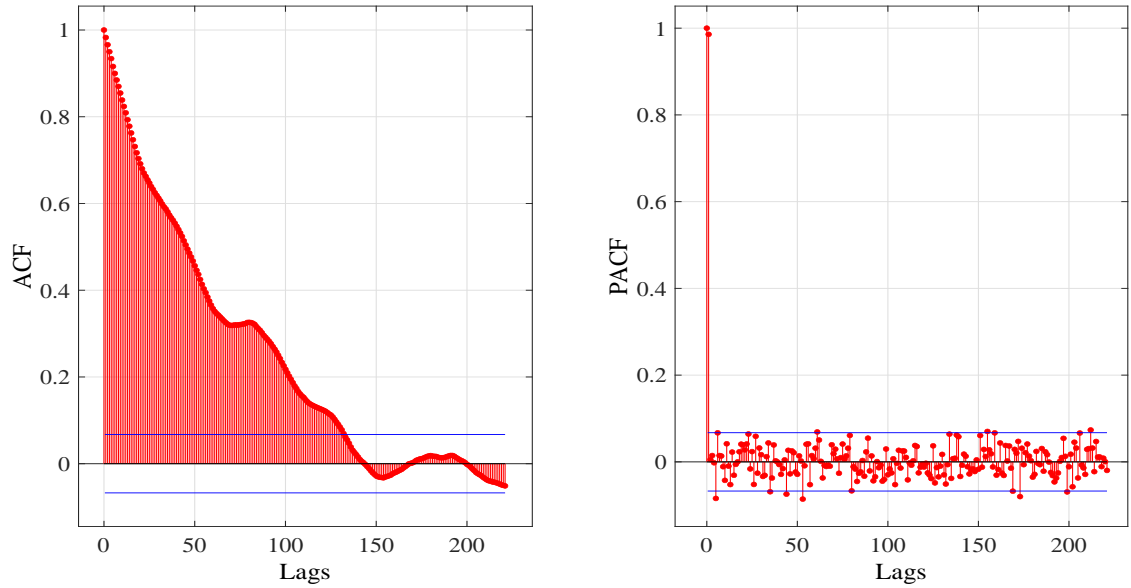}
\caption{The ACF (left panel) and PACF (right panel) of the time series of the simulated QPPs presented in Figure \ref{Fig2}.}
\label{Fig9}
\end{figure}

\begin{figure}
\centering
\includegraphics[width=18cm,height=8.5cm]{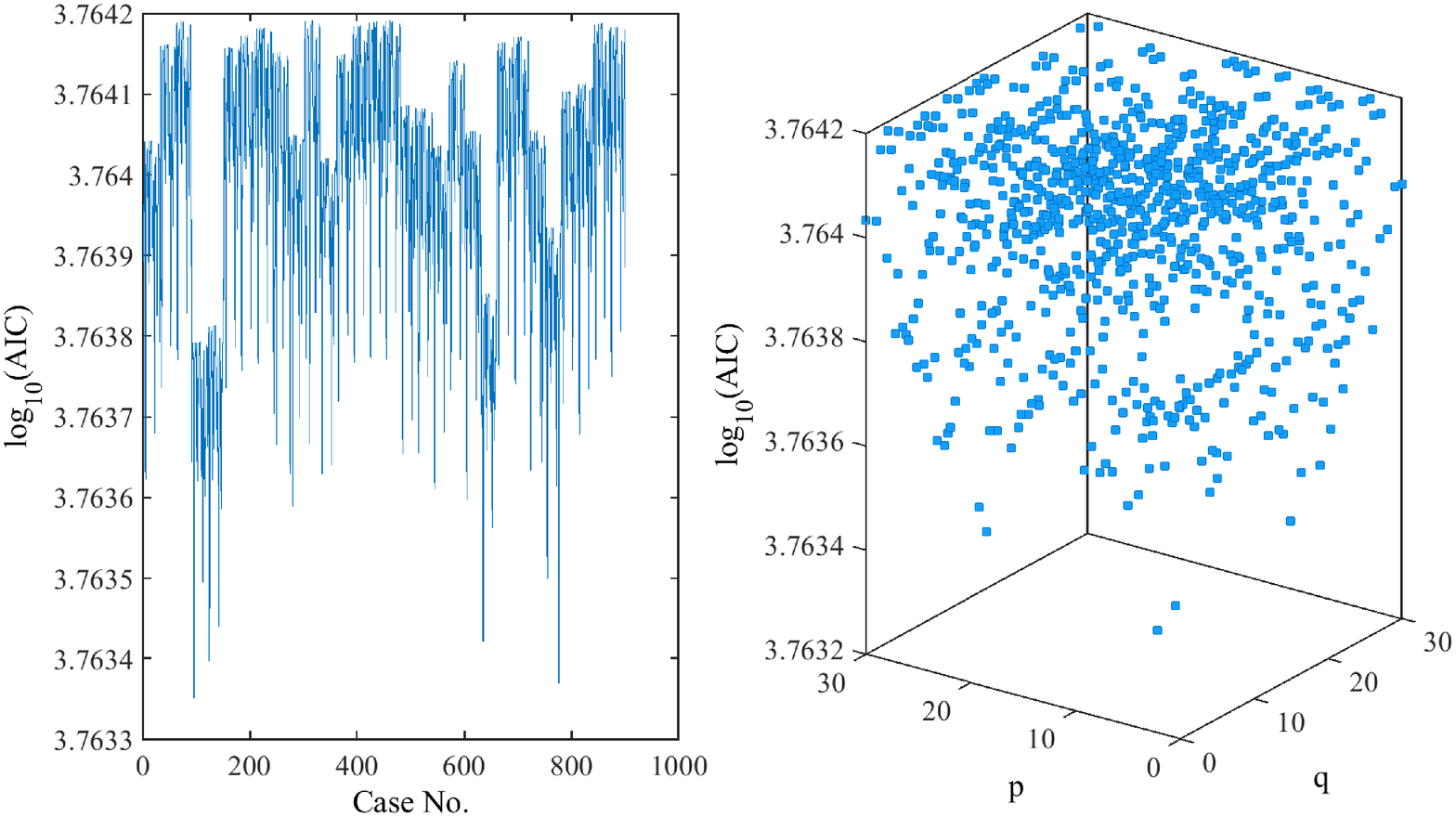}
\caption{The AICs of $900$ ARIMA models applied on the simulated QPPs of Figure \ref{Fig2} with $p,q=1..30,$ and $d=1$.}
\label{Fig10}
\end{figure}

\begin{figure}
\centering
\includegraphics[width=8.5cm,height=7cm]{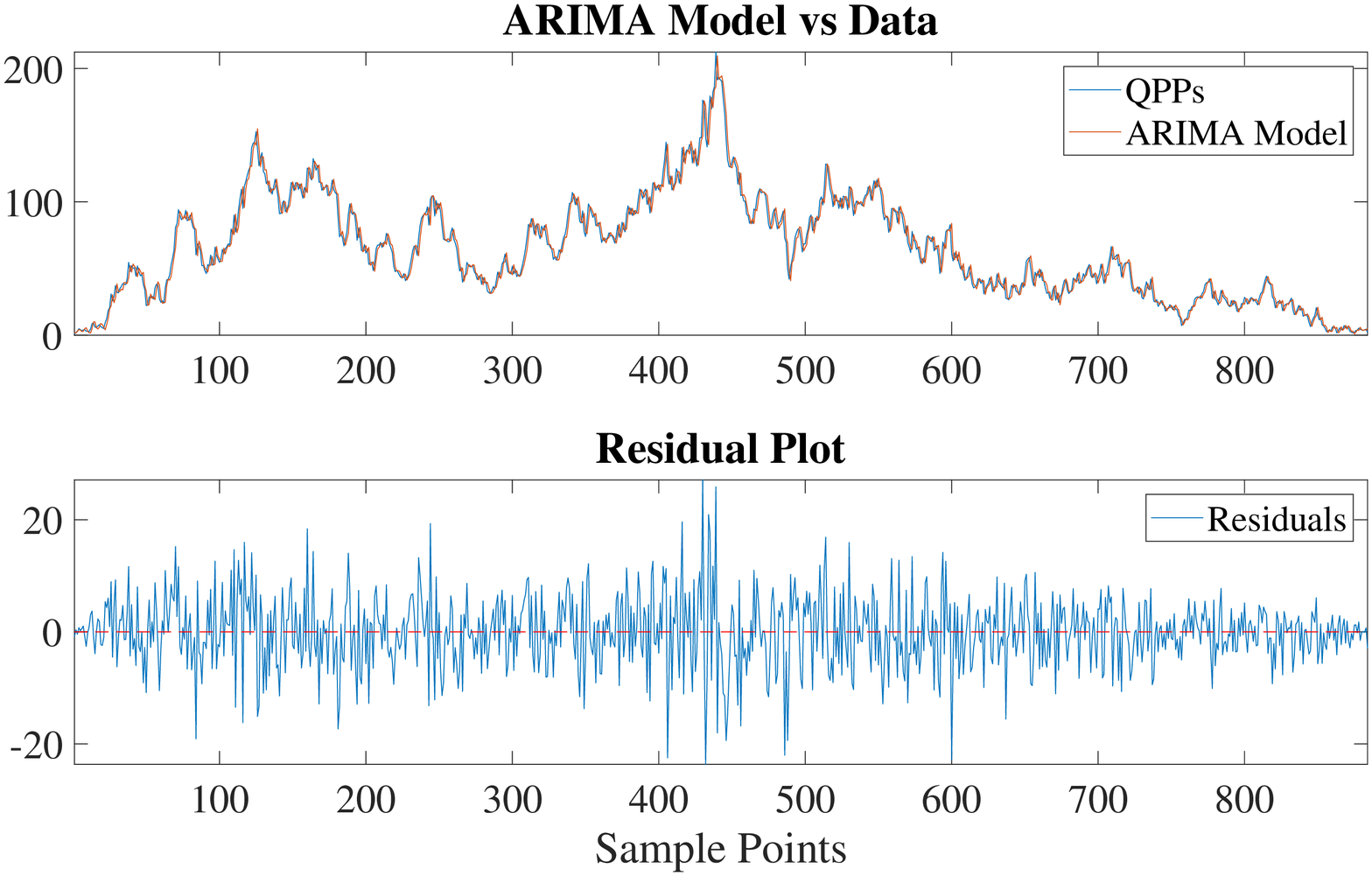}
\includegraphics[width=8.5cm,height=7cm]{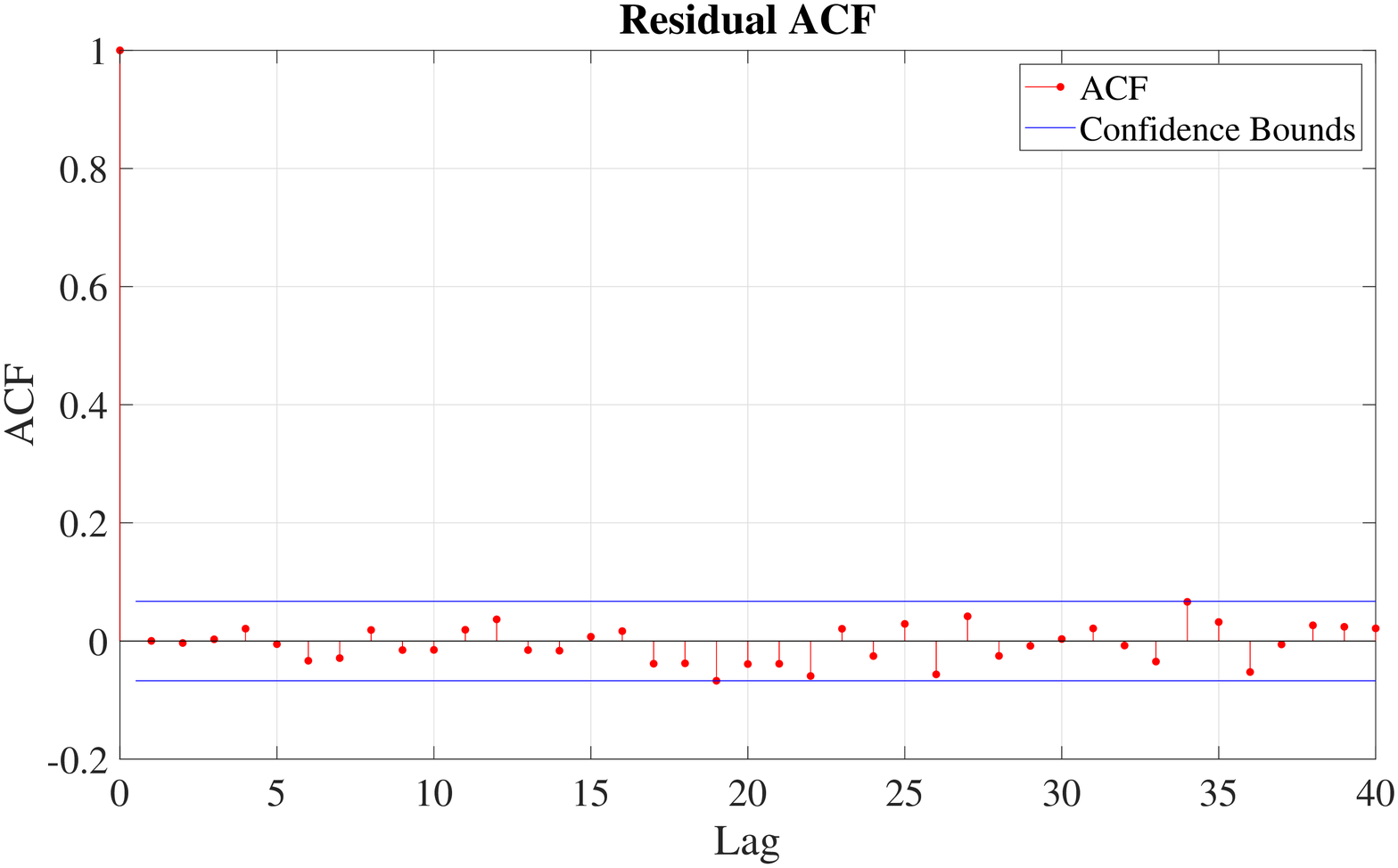}
\caption{The simulated QPPs of Figure \ref{Fig2} vs the ARIMA(4,1,5) model (top left panel), the obtained residual (bottom left panel), and the ACF of the residuals (right panel).}
\label{Fig11}
\end{figure}

Figure \ref{Fig9} illustrates the autocorrelation function (ACF) and partial autocorrelation function (PACF) of the simulated QPPs. The lag number and the significant range are considered a quarter of the sample points and $ \pm 2\sigma, $ respectively. As seen in the figure, the ACF tails off gradually which is consistent with the existing trend in the time series. This indicates that a non-zero difference parameter is required for modeling. Moreover, a periodic-like pattern is observed in the ACF. The PACF also spikes strongly at lag 1. The autocorrelations provide some intuition about the order of parameters. However, the preferred values are those minimizing the Akaiki information criterion (AIC). AIC estimates the performance of an applied model on a given data sample. Further to this criterion, the residuals should be stationary and normally distributed. Figure \ref{Fig10} displays AICs for $900$ examined ARIMA models. The minimum AIC corresponds to $p=4, d=1, $ and $q=5$. Strong negative spikes are observed for models with $d=2$ as a manifestation of overdifferencing. Figure \ref{Fig11} shows the simulated time series together with the ARIMA(4,1,5) model, and the residuals (left panels).

Even though the AIC has determined a decent model, an explicit investigation of residuals is still required to avoid possible mis-specifications. Besides stationarity, the residuals of a convenient ARIMA-type model should follow a Gaussian distribution with no correlations. In order to check whether these conditions are met, various tests are available e.g., Ljung-Box Q-test for autocorrelations, KS-test, KPSS test, Dickey–Fuller test, and etc. \citep[see the Econometrics Toolbox User's Guide,][for a detailed review]{matlab}. We performed a detrended fluctuation analysis (DFA) on the residuals and obtained the Hurst exponent equal to $ 0.48 \pm 0.01 $. We also investigated the existence of a unit root applying the mentioned tests and found that the residuals have similar characteristics to a white noise. The ACF of the residuals is shown in Figure \ref{Fig11} right panel.

Generally, quasi-periodic patterns are universal features that arise due to an underlying $1/\nu-$type process \citep{feigelson2012modern}. Such long-memory processes might be described with FARIMA models. The difference parameter of a stationary and invertible fractional ARIMA model lies in the range of $ -1/2 < d_{f} < 1/2 $, where $ d_{f} $ corresponds to the long-term dependency in the subject time series \citep{beran2013}. There are several methods to measure the difference parameter of a FARIMA model such as performing a DFA (as $H = d_{f} + 1/2$), fitting a power law to either the gradual decay of ACF ($ACF(l) \propto l^{2d_{f}-1} $) or the Fourier PSD ($ \mathcal{P}(\nu) \propto \lvert \nu \lvert ^{-2d_{f}}$), or even performing a discrete wavelet transformation \citep{Barkoulas1997, reisen2001}. On the downside, different methods might not necessarily converge to the same result \cite[see][section 11.9]{feigelson2012modern}.

Applying a DFA, the Hurst exponent of the simulated QPPs is obtained $ 1.35 \pm 0.01 $ which means that the time series is non-stationary, and leads to $ d_{f} = 0.85 $. Furthermore, we performed a power-law fit on the ACF of the QPPs and found $ d_{f} = 0.75 $. The obtained fractional differences imply that even though the process is mean reverting, the variance is infinite \citep{granger1980}. \cite{beran2013} discussed cases for which both the fractional and non-fractional difference parameters are required, namely, FARIMA($p,d_{\rm total},q$) models with $ d_{\rm total} = d + d_{f}$. One possible approach is to consider e.g., $d=1$ and take the difference of the time series through $(1-B)S_{t}$, then, compute $d_{f}$. Following this perspective, we obtain $ d_{f} = -0.03$ which indicates a relatively weak long-term dependency.

\begin{figure}
\centering
\includegraphics[width=18cm,height=12cm]{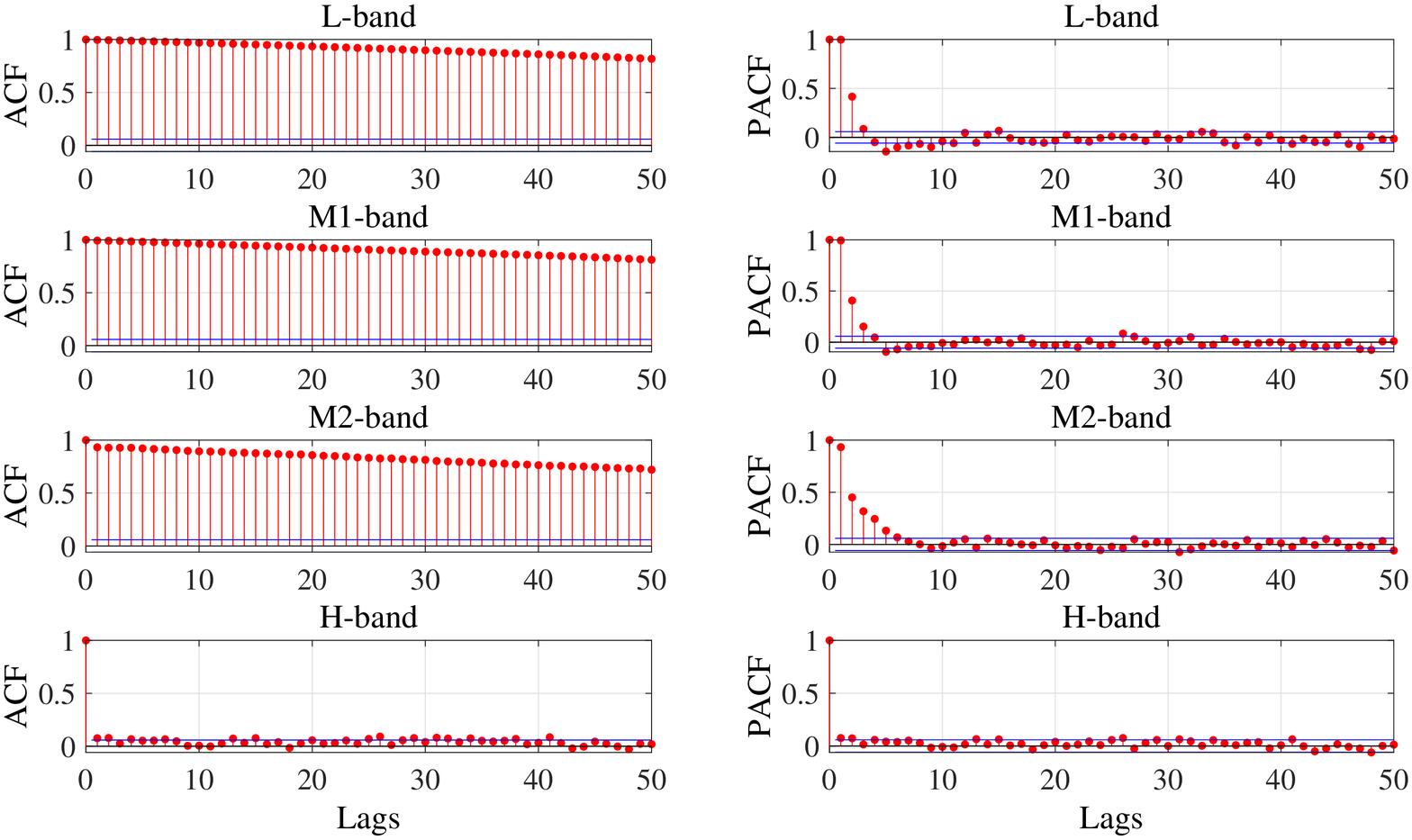}
\caption{The ACFs and PACFs of the HXR emissions of the solar flare displayed in Figure \ref{Fig3}.}
\label{Fig12}
\end{figure}

\begin{figure}
\centering
\includegraphics[width=8.75cm,height=6cm]{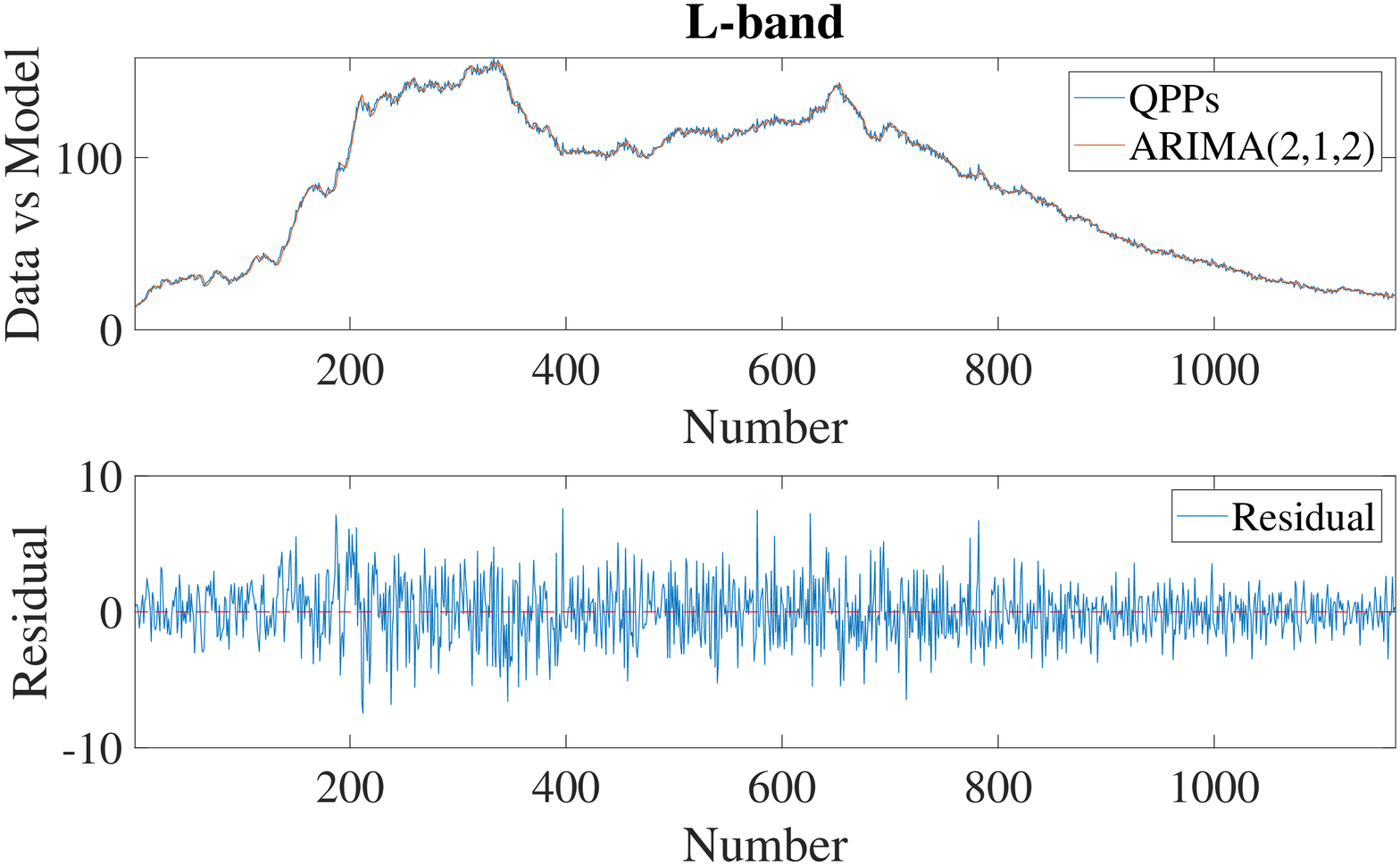}
\includegraphics[width=8.75cm,height=6cm]{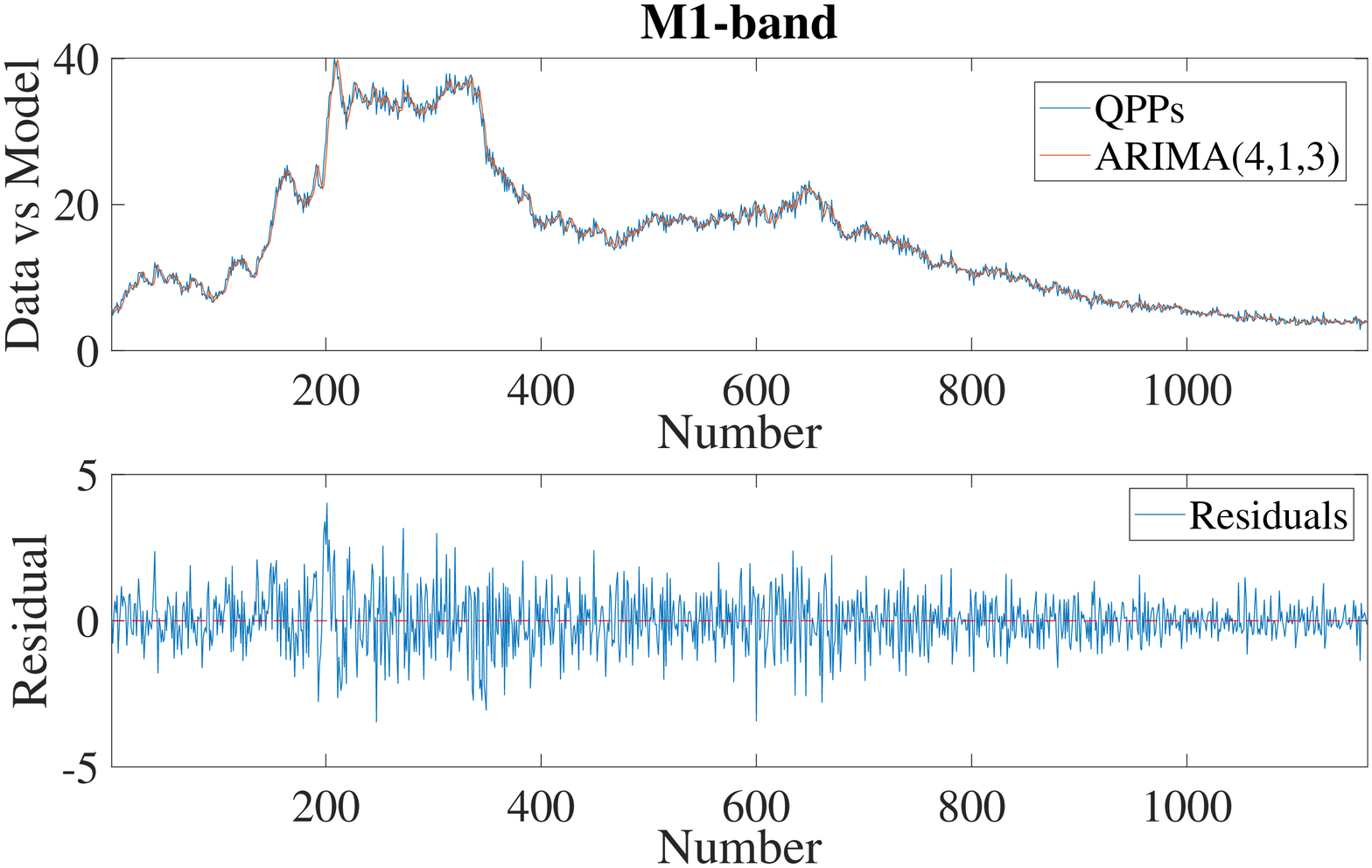}
\includegraphics[width=8.75cm,height=6cm]{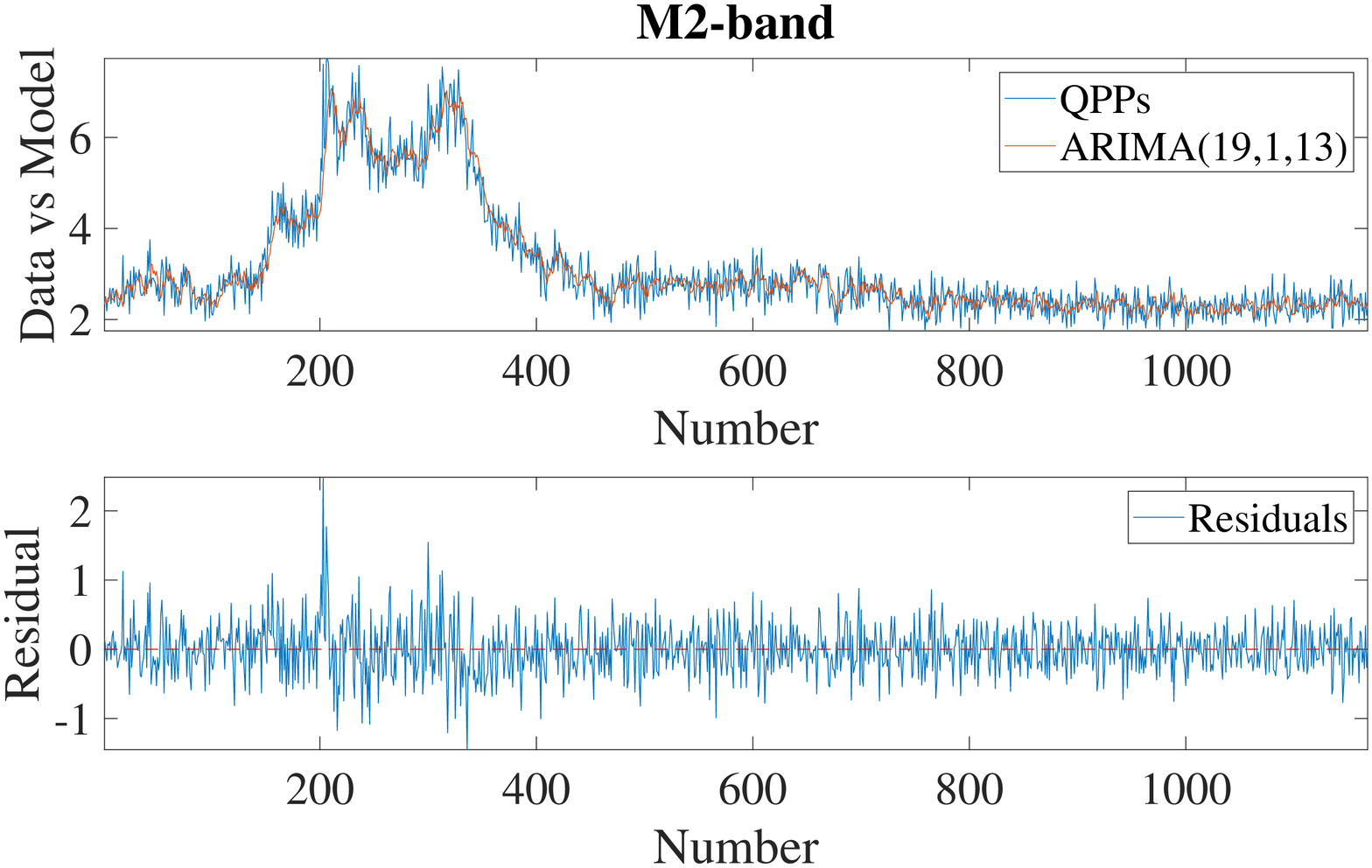}
\includegraphics[width=8.75cm,height=6cm]{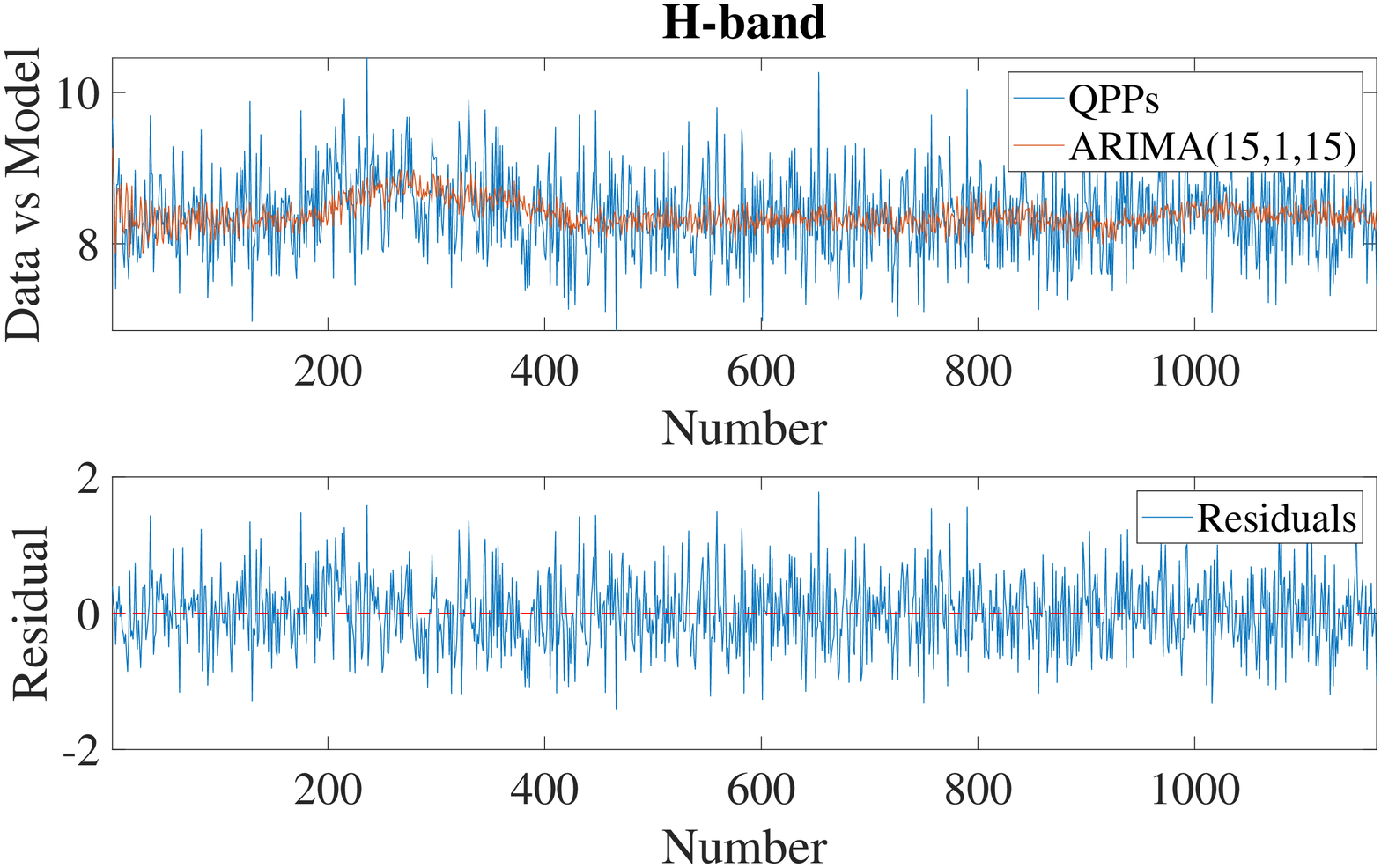}
\caption{The HXR emissions of the solar flare of Figure \ref{Fig3} vs the selected ARIMA models and relevant residuals in four spectral bands of L, M1, M2, and H.}
\label{Fig14}
\end{figure}

We have also performed a similar analysis on the HXR emissions of the solar flare of Figure \ref{Fig3}. The ACFs and PACFs relevant to each pass band are shown in Figure \ref{Fig12}. Various choices of ARIMA parameters ($ p,q = 1..20, $ and $d=1,2$) are examined and their AICs are measured. The minimum AICs of the L, M1, M2, and H bands are achieved for ARIMA$(2,1,2)$, ARIMA$(4,1,3)$, ARIMA$(19,1,13)$ and ARIMA$(15,1,15)$, respectively. However, the residuals do not follow a Gaussian distribution in contrast with the initial assumption of Equation (\ref{eqapp3}). Such violation might impact the maximum likelihood estimations performed for AIC calculation and model selection. Nonetheless, the obtained parameters might still be correct as the regression models could practically overcome the non-normality of residuals in many cases \citep[see e.g.,][]{knief2021violating}. Figure \ref{Fig14} displays the registered emissions, ARIMA models and the residuals for each spectral band.

It seems that ARIMA models could conveniently characterize the time series of simulated flaring events. For the observational time series, more study is required to evaluate the influence of initial assumptions on model selection. In the next step, we would like to extend the study by considering a none-Gaussian noise in ARIMA-type models to adequately investigate their practicality in flare studies.

\section{Conclusion}\label{sec:con}
The scale-free nature of solar flare energies inspired the application of SOC models to investigate these phenomena. Despite the development of numerous MHD simulation models, less attention has been paid to the pre-existing CA models and their capabilities in simulating QPPs. In the present study, we took the idea of SOC together with the self-oscillatory processes (load/unload mechanism) into consideration and discussed that CA models could technically generate QPPs due to their intrinsic ability in reproducing repetitive reconnection regimes. Then, we reappraised the well-known CA avalanche model of \citeauthor{LH1991} and investigated the productivity of this model in evaluating QPPs' characteristics.

Applying the modified Lu \& Hamilton model, we simulated $21,070$ flaring events. We obtained that nearly $4 \%$ of these events last over $50$ seconds, scaled with the HXT temporal resolution, which mostly includes the C-class flares and more energetic explosions (some B-class flares are also included in this list). The QPPs are found in $70 \%$ of these flares ($565$ out of $813$). However, less than $1 \%$ of small-scale events ($D \le 50$ seconds) exhibited quasi-periodic patterns. We applied the Lomb-Scargle periodogram to study the quasi-periodic patterns. The distribution of extracted periodicities follows a lognormal distribution with a global maximum of $29.29 \pm 0.67$ seconds which illustrates the most probable period for the simulated QPPs. The obtained lognormal behavior indicates the presence of a multiplicative mechanism (typifying the effect of independent varying parameters e.g., driving mechanism, number of unstable sites, released energies, etc.) due to which the system evolves. However, the stochastic and scale-free nature is preserved in the memory of the time series as an essential characteristic of the observational/simulated flaring events.

We observed that the CA models could practically produce a wide range of periodicities for QPPs. Moreover, the presence of multiple periods is observed in nearly $50 \%$ of simulated QPPs. According to our results, although there is a clear dependency between the QPPs' periods and the host flare durations, other covariates might also be involved that affect their relation. We obtained that the applied CA model adequately accomplishes simulating flares, QPPs, and their statistics. Moreover, we examined ARIMA-type models on the time series of observational and simulated QPPs. We observed that ARIMA models could describe the subject QPPs. However, further studies are required to address their utility in modeling flares and accompanying QPPs.

\textbf{Acknowledgments:} We acknowledge the use of data from the YLA. The authors gratefully thank Dr. Aki Takeda and Dr. Keiji Yoshimura from the YLA team for their kind response and preparation of the HXR time profiles. N.F. express her gratitude to the Iran National Science Foundation (INSF) for supporting this research under grant No. 99012824. The authors also gratefully acknowledge the anonymous Reviewer and statistics Editor for their constructive suggestions.

\newpage

\newpage
\clearpage
\bibliography{ref_2022.bib}
\end{document}